\pdfoutput=1

\documentclass[iop]{emulateapj}

\usepackage{natbib,epsfig,siunitx,url,graphicx,amsmath,footnote,float,xcolor,setspace,cases}
\usepackage[colorlinks=true,linkcolor=blue,citecolor=blue]{hyperref}

\begin{document}

\submitted{Icarus; accepted}

\title{Mars' Growth Stunted by an Early Giant Planet Instability}

\author{Matthew S. Clement\altaffilmark{1,*}, Nathan A. Kaib\altaffilmark{1}, Sean N. Raymond\altaffilmark{2}, \& Kevin J. Walsh\altaffilmark{3}}

\altaffiltext{1}{HL Dodge Department of Physics Astronomy, University of Oklahoma, Norman, OK 73019, USA}
\altaffiltext{2}{Laboratoire d’Astrophysique de Bordeaux, Univ. Bordeaux, CNRS, B18N, allée Geoffroy Saint-Hilaire, 33615 Pessac, France}
\altaffiltext{3}{Southwest Research Institute, 1050 Walnut St. Suite 300, Boulder, CO 80302, USA}
\altaffiltext{*}{corresponding author email: matt.clement@ou.edu}

\setcounter{footnote}{0}
\begin{abstract}
	Many dynamical aspects of the solar system can be explained by the outer planets experiencing a period of orbital instability sometimes called the Nice Model.  Though often correlated with a perceived delayed spike in the lunar cratering record known as the Late Heavy Bombardment (LHB), recent work suggests that this event may have occurred much earlier; perhaps during the epoch of terrestrial planet formation.  While current simulations of terrestrial accretion can reproduce many observed qualities of the solar system, replicating the small mass of Mars requires modification to standard planet formation models.  Here we use 800 dynamical simulations to show that an early instability in the outer solar system strongly influences terrestrial planet formation and regularly yields properly sized Mars analogs.  Our most successful outcomes occur when the terrestrial planets evolve an additional 1-10 million years (Myr) following the dispersal of the gas disk, before the onset of the giant planet instability.   In these simulations, accretion has begun in the Mars region before the instability, but the dynamical perturbation induced by the giant planets' scattering removes large embryos from Mars' vicinity.  Large embryos are either ejected or scattered inward toward Earth and Venus (in some cases to deliver water), and Mars is left behind as a stranded embryo.  An early giant planet instability can thus replicate both the inner and outer solar system in a single model.
\end{abstract}

\section{Introduction}

It is widely understood that the evolution of the solar system's giant planets play the most important role in shaping the dynamical system of bodies we observe today.  When the outer planets interact with an exterior disk of bodies, Saturn, Uranus and Neptune tend to scatter objects inward \citep{fer84}.  To conserve angular momentum through this process, the orbits of these planets move outward over time \citep{hahn99,gomes03}.  Thus, as the young solar system evolved, the three most distant planets' orbits moved out while Jupiter (which is more likely to eject small bodies from the system) moved in.  To explain the excitation of Pluto's resonant orbit with Neptune, \citet{malhotra93} proposed that Uranus and Neptune must have undergone significant orbital migration prior to arriving at their present semi-major axes.  \citet{malhotra95} later expanded upon this idea to explain the full resonant structure of the Kuiper belt. In the same manner, an orbital instability in the outer solar system can successfully excite Kuiper belt eccentricities and inclinations, while simultaneously moving the giant planets to their present semi-major axes via planet-planet scattering followed by dynamical friction \citep{thommes99}.

These ideas culminated in the eventual hypothesis that, as the giant planets orbits diverged after their formation, Jupiter and Saturn's orbits would have crossed a mutual 2:1 Mean Motion Resonance (MMR).  Known as the Nice (as in Nice, France) Model \citep{Tsi05,gomes05,mor05}, this resonant configuration of the two most massive planets causes a solar system-wide instability, which has been shown to reproduce many peculiar dynamical traits of the solar system.  This hypothesis has subsequently explained the overall structure of the Kuiper belt  \citep{levison08,nesvorny15a,nesvorny15b}, the capture of trojan satellites by Jupiter \citep{mor05,nesvorny13}, the orbital architecture of the asteroid belt \citep{roig15}, and the giant planets' irregular satellites, including Triton \citep{nesvory07}.

The Nice Model itself has changed significantly since its introduction.  In order to  keep the orbits of the terrestrial planets dynamically cold (low eccentricities and inclinations), \citet{bras09} proposed that Jupiter ``jump'' over its 2:1 MMR with Saturn, rather than migrate smoothly through it \citep{morby09,morby10}.  Otherwise, the terrestrial planets were routinely excited to the point where they were ejected or collided with one another in simulations.  The probability of producing successful jumps in these simulations is greatly increased when an extra primordial ice giant was added to the model \citep{nesvorny11,batbro12}.  In successful simulations, the ejection of an additional ice giant rapidly forces Jupiter and Saturn across the 2:1 MMR.  Furthermore, hydrodynamical simulations \citep{snellgrove01,np1} show that a resonant chain of giant planets is likely to emerge from the dissipating gaseous circumstellar disk, with Jupiter and Saturn locked in an initial 3:2 MMR resonance \citep{masset01,morbidelli07,pierens08}. This configuration can produce the same results as the 2:1 MMR crossing model \citep{morby07}.  In this scenario, the instability ensues when two giant planets fall out of their mutual resonant configuration.  Such an evolutionary scheme seems consistent with the number of resonant giant exoplanets discovered (eg: HD 60532b, GJ 876b, HD 45364b, HD 27894, Kepler 223 and HR 8799 \citep{holman10,fab10,rivera10,laskar15,mills16,Trifonov17}).

Despite the fact that 5 and 6 primordial giant planet configurations are quite successful at reproducing the architecture of the outer solar system \citep{nesvorny12}, delaying the instability $\sim$400 Myr to coincide with the lunar cataclysm \citep{gomes05} still proves problematic for the terrestrial planets.   Indeed, \citet{kaibcham16} find only a $\sim1\%$ chance that the terrestrial planets' orbits and the giant planets' orbits are reproduced simultaneously.  Even in systems with an ideal ``jump,'' the eccentricity excitation of Jupiter and Saturn can bleed to the terrestrial planets via stochastic diffusion, leading to the over-excitation or ejection of one or more inner planets \citep{agnorlin12,bras13,roig15}.  It should be noted, however, that Mercury's uniquely excited orbit (largest mean eccentricity and inclination of the planets) may be explained by a giant planet instability \citep{roig16}.  Nevertheless, the chances of the entire  solar system emerging from a late instability in a configuration roughly resembling its modern architecture are very low \citep{kaibcham16}.  This suggests that the instability is more likely to have not occurred in conjunction with the LHB, but rather before the terrestrial planets had fully formed.  Fortunately, many of the dynamical constraints on the problem are fairly impartial to whether the instability happened early or late \citep{morb18}.  The Kuiper belt's orbital structure  \citep{levison08,nesvorny15a,nesvorny15b}, Jupiter's Trojans \citep{mor05,nesvorny13}, Ganymede and Callisto's different differentiation states \citep{barr10} and the capture of irregular satellites in the outer solar system \citep{nesvory07} are still explained well regardless of the specific timing of the Nice Model instability.  In addition to perhaps ensuring the survivability of the terrestrial system, there are several other compelling reasons to investigate an early instability:

$\textbf{1. Uncertainties in Disk Properties:}$  Since the introduction of the Nice Model, simplifying assumptions of the unknown properties of the primordial Kuiper belt have provided initial conditions for N-body simulations.  The actual timing of the instability is highly sensitive to the particular disk structure selected \citep{gomes05}.  Furthermore, numerical studies must approximate the complex disk structure with a small number of bodies in order to optimize the computational cost of simulations.  In fact, most N-body simulations do not account for the affects of disk self gravity \citep{nesvorny12}.  When the giant planets are embedded in a disk of gravitationally self-interacting particles using a graphics processing unit (GPU) to perform calculations in parallel and accelerate simulations \citep{genga}, instabilities typically occur far earlier than what is required for a late instability (Quarles $\&$ Kaib, in prep).

$\textbf{2. Highly Siderophile Elements (HSE):}$ A late instability (the LHB) was originally favored because of the small mass accreted by the Moon relative to the Earth after the Moon-forming impact (for a review of these ideas see \citet{morb12}).  The HSE record from lunar samples indicates that the Earth accreted almost 1200 times more material, despite the fact that its geometric cross-section is only about 20 times that of the Moon \citep{walker04,day07,walker09}.  Thus, the flux of objects impacting the young Earth would have had a very top-heavy size distribution (\citet{bottke10}, however \citet{minton15} showed that the pre-bombardment impactor size distribution may not be as steep as originally assumed).  This distribution of impactors is greatly dissimilar from what is observed today, and favors the occurrence of a LHB.  New results, however, indicate that the HSE disparity is actually a result of iron and sulfur segregation in the Moon's primordial magma ocean causing HSEs to drag towards the core long after the moon-forming impact \citep{rubie16}.  Because the crystallization of the lunar magma took far longer than on Earth, a large disparity between the HSE records is expected \citep{morb18}.

$\textbf{3. Updated Impact Data:}$  The LHB hypothesis gained significant momentum when none of the lunar impactites returned by the Apollo missions were older than ~3.9 Gyr \citep{tera74,zellner17}.  However, recent $^{40}Ar/^{39}Ar$ age measurements of melt clasts in Lunar meteorites are inconsistent with the U/Pb dates determined in the 1970s \citep{fernandes00,chapman07,boehnke16}.   These new dates cover a broader range of lunar ages; and thus imply a smoother decline of the Moon's cratering rate.  Furthermore, new high-resolution images from the Lunar reconnaissance Orbiter (LRO) and the GRAIL spacecraft have significantly increased the number of old ($>$3.9 Gyr) crater basins used in crater counting \citep{spudis11,fassett12}.  For example, samples returned by Apollo 17 that were originally assumed to be from the impactor that formed the Serenitatis basin are likely contaminated by ejecta from the Irbrium basin \citep{spudis11}.  Because the Serenitatis basin is highly marred by young craters and ring structures, it is likely older than ~4 Gyr, and the Apollo samples are merely remnants of the 3.9 Gyr Irbrium event.

Here we build upon the hypothesis of an early instability by systematically investigating the effects of the Nice Model occurring during the process of terrestrial planetary formation.  Since advances in algorithms substantially decreased the computational cost of N-body integrators in the 1990s \citep{wisdomholman,duncan98,chambers99}, many papers have been dedicated to modeling the late stages (giant impact phase) of terrestrial planetary formation.  Observations of proto-stellar disks \citep{haisch01,pasucci09} suggest that free gas disappears far quicker than the timescale radioactive dating indicates it took the terrestrial planets to form \citep{halliday08,kleine09}.  Because the outer planets must clearly form first, the presence of Jupiter is supremely important when modeling the formation of the inner planets \citep{wetherill96,chambers_cassen02,levison_agnor03}.  Early N-body integrations of planet formation in the inner solar system in 3 dimensions from a disk of planetary embryos and a uniformly distributed sea of planetesimals reproduced the general orbital spacing of our 4 terrestrial planets \citep{chambers98,chambers01}.  However, these efforts systematically failed to produce an excited asteroid belt and 4 dynamically cold planets with the correct mass ratios (Mercury and Mars are $\sim5\%$ and $\sim10\%$ the mass of Earth respectively).

\begin{table*}
\centering
\caption{Giant Planet Initial Conditions: The columns are: (1) the name of the simulation set, (2) the number of giant planets, (3) the mass of the planetesimal disk exterior to the giant planets, (4) the distance between the outermost ice giant and the planetesimal disk’s inner edge, (5) the semi-major axis of the outermost ice giant (commonly referred to as Neptune, however not necessarily the planet which completes the simulation at Neptune's present orbit), (6) the resonant configuration of the giant planets starting with the Jupiter/Saturn resonance, and (7) the masses of the ice giants from inside to outside.}
\begin{tabular}{c c c c c c c c}
\hline
Name  & $N_{Pln}$ &  $M_{disk}$ & $\delta$r & $r_{out}$ & $a_{nep}$ & Resonance Chain & $M_{ice}$\\
& & ($M_{\oplus}$) & (au) & (au) & (au) & & ($M_{\oplus}$)\\
\hline
n1 & 5 & 35 & 1.5 & 30 & 17.4 & 3:2,3:2,3:2,3:2 & 16,16,16 \\
n2 & 6 & 20 & 1.0 & 30 & 20.6 & 3:2,4:3,3:2,3:2,3:2 & 8,8,16,16 \\
\hline
\end{tabular}
\label{table:gp}
\end{table*}

Numerous subsequent authors approached these problems using various methods and initial conditions.  By accounting for the dynamical friction of small planetesimals, \citet{obrien06} and \citet{ray06} more consistently replicated the low eccentricities of the terrestrial planets.  However, the so-called ``Small Mars Problem'' proved to be a systematic short-coming of N-body accretion models \citep{wetherill91,ray09a}.  The vast majority of simulations produce Mars analogs roughly the same mass as Earth and Venus; a full order of magnitude too large \citep{morishma10}.  However, Mars' mass consistently stayed low when using a configuration of Jupiter and Saturn with present day mutual inclination and eccentricities twice their modern values \citep{ray09a}.  Because planet-disk interactions systematically damp the eccentricities of growing gas planets \citep{papaloizou00,tanaka04}, this result presented the problem of requiring a mechanism to adequately excite the orbits of the giant planets prior to terrestrial planetary formation.  \citet{hansen09} then demonstrated that a small Mars could be formed if the initial disk of planetesimals was confined to a narrow annulus between 0.7-1.0 au (interestingly, a narrow annulus might also explain the orbital distribution of silicate rich S-type asteroids \citep{ray17sci}).  The ``Grand Tack'' hypothesis provides an interesting mechanism to create these conditions whereby a still-forming Jupiter migrates inward and subsequently ``tacks'' backward once it falls into resonance with Saturn \citep{walsh11,brasser16}.  When Jupiter ``tacks'' at the correct location, the disk of planetesimals in the still-forming inner solar system is truncated at 1.0 au, roughly replicating an annulus (for a critical review of the Grand Tack consult \citet{raymorb14}).  

Another potential solution to the small Mars problem is local depletion of the outer disk \citep{iz14,izidoro15}.  However, systems in these studies that placed Jupiter and Saturn on more realistic initially circular orbits failed to produce a small Mars.  Mars' formation could have also been affected by a secular resonance with Jupiter sweeping across the inner solar system as the gaseous disk depletes \citep{thommes08,bromley17}.  The degree to which this process induces a dynamical shake-up of material in the vicinity of the forming Mars and asteroid belt is strongly tied to speed of the resonance sweeping.  Furthermore, it is possible that Mars' peculiar mass is simply the result of a low probability event.  Indeed, there is a low, but non-negligible probability of forming a small Mars when using standard initial conditions and assuming no prior depletion of the disk \citep{fischer14}.  However, on closer inspection, many of the ``successful'' systems in \citet{fischer14} are poor solar system analogs for other reasons (such as an extra large planet in the asteroid belt \citep{jacobson15}).  Additionally, the large masses of Mars analogs produced in N-body integrations could be a consequence of the simplifications and assumptions made by such simulations.  The process of how planetesimals form out of small, pebble to meter sized bodies is still an active field of research.  Reevaluating the initial conditions used by N-body accretion models of the giant impact phase may potentially shed light on the origin of Mars' small mass. \citet{kenyon06} considered this problem using a multi-annulus coagulation code to grow kilometer scale planetesimals.  Subsequent authors \citep{levison15,draz16,ray17sci} modeling the accretion of meter sized objects found that dust growth and drift cause solids in the inner disk to be redistributed in to a much steeper radial profile.  Finally, multiple studies have demonstrated that more realistic, erosive collisions can significantly reduce the mass of embryos during the late stages of terrestrial planet accretion \citep{kokubo10,kobayashi13,chambers13}.

Here we investigate an alternative scenario wherein the still forming inner planets are subjected to a Nice Model instability.  Though the effect of the Nice Model on the fully formed terrestrial planets is well studied \citep{bras09,agnorlin12,bras13,kaibcham16,roig16}, no investigation to date has performed direct numerical simulations of the effect of the Nice Model instability on the still forming terrestrial planets.  Furthermore, our work is motivated by simulations from \citet{lykawaka13}.  The authors found that a low massed Mars could be formed when Jupiter and Saturn (with enhanced eccentricities) were artificially migrated across their mutual 2:1 MMR using fictitious forces.  A downfall of this scenario, however, is that it overexcites the orbits of the forming terrestrial planets.  Additionally, previous authors \citep{walshmorb11} have investigated whether smooth migration of Jupiter and Saturn (as opposed to the ``jumping Jupiter'' model) could produce a small Mars, however the speed of the process was found to be too slow with respect to Mars' formation timescale (1-10 Myr; \citet{Dauphas11}).  The work of \citet{walshmorb11} is perhaps the most similar to that of this paper.  However, our work differs greatly in that we consider full instabilities directly, rather than modeling terrestrial evolution while migrating the giant planets with artificial forces.  It should also be noted that we do not include the ``Grand Tack'' hypothesis in our study (in section 5 we argue that it is potentially compatible with the scenario we investigate).

\section{Materials and Methods}

Because the parameter space of possible giant planet configurations (number of planets, resonant configuration, planet spacing, disk mass and disk spacing) emerging from the primordial gas disk is substantial, as a starting point for this work we take two of the most successful five and six giant planet configurations from \citet{nesvorny12}\footnote[1]{It should be noted that, with the additional constraint of requiring that Neptune migrate to $\sim$28 au prior to the onset of the instability, a 3:2,3:2,2:1,3:2 resonant configuration for 5 planet scenarios is more effective \citep{deienno17}.  Because we are mostly interested in studying the excitation of the inner solar system, which is largely unaffected by the particular migration of Neptune, our results should be relatively independent of the particular ice giant resonant configuration selected.}.  Both begin with Jupiter and Saturn in their 3:2 MMR.  Though placing Jupiter and Saturn in an initial 2:1 configuration on circular orbits can be highly successful at replicating the correct planetary spacing of the outer solar system, only $\sim$0.2$\%$ such simulations sufficiently excite Jupiter's eccentricity \citep{nesvorny12}.  For this reason we focus on the scenario where Jupiter and Saturn emerge from the gas disk locked in a 3:2 MMR.  Furthermore, \citet{nesvorny12} found advantages and disadvantages which are mutually exclusive to both the five and six planet cases.  Thus, for completeness we select one five and one six planet setup for our study.  We summarize our chosen sets of giant planet initial conditions in table \ref{table:gp}.

To create a mutual resonant chain of gas giants, we apply an external force to the giant planets which mimics the effects of a gas disk by modifying the equations of motion with forced migration ($\dot{a}$) and eccentricity damping ($\dot{e}$) terms \citep{lee02}.  Though the precise underlying physics of the interaction between forming giant planets and a gas disk is not fully understood, this method is employed by many authors studying both the solar system, and observed resonant exoplanets because it consistently produces stable resonant chains \citep{matsu10,beauge12}.  The exact functional forms of $\dot{a}$ and $\dot{e}$ depend on the timescale and overall distance of migration desired.    We utilize a form of $\dot{a}=ka$ and $\dot{e}=ke/100$ \citep{bat12}, where the constant k is adjusted to achieve a migration timescale $\tau_{mig}\sim$.1-1 Myr.  Because the actual migration rate is a complex function of the properties of the gas disk and relative masses of the planets, achieving a specific migration timescale is of less importance than placing the resonant planets on the proper orbits \citep{kley12,baruteau14}

\begin{figure}
\centering
\includegraphics[width=.5\textwidth]{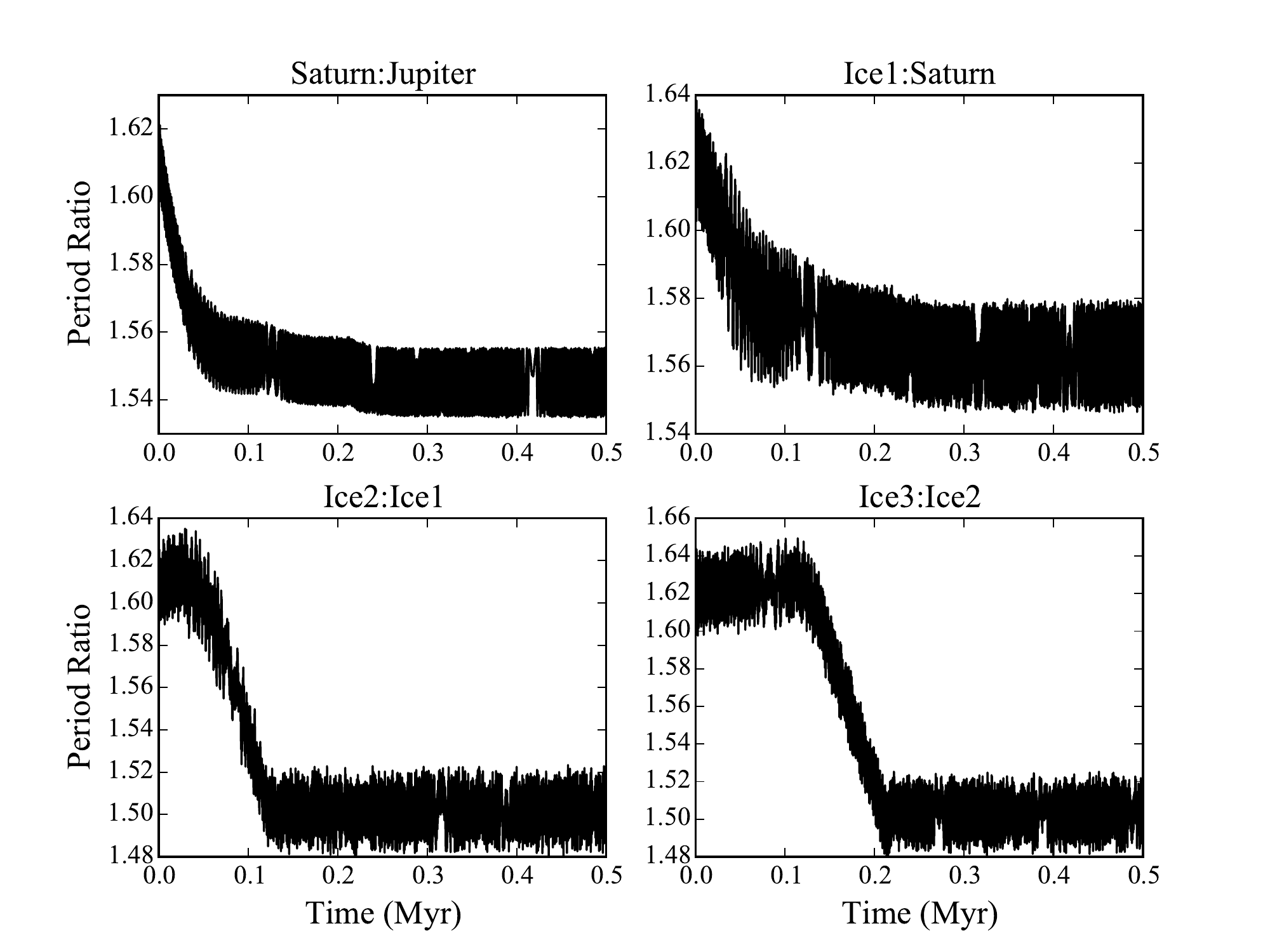}
\qquad
\includegraphics[width=.5\textwidth]{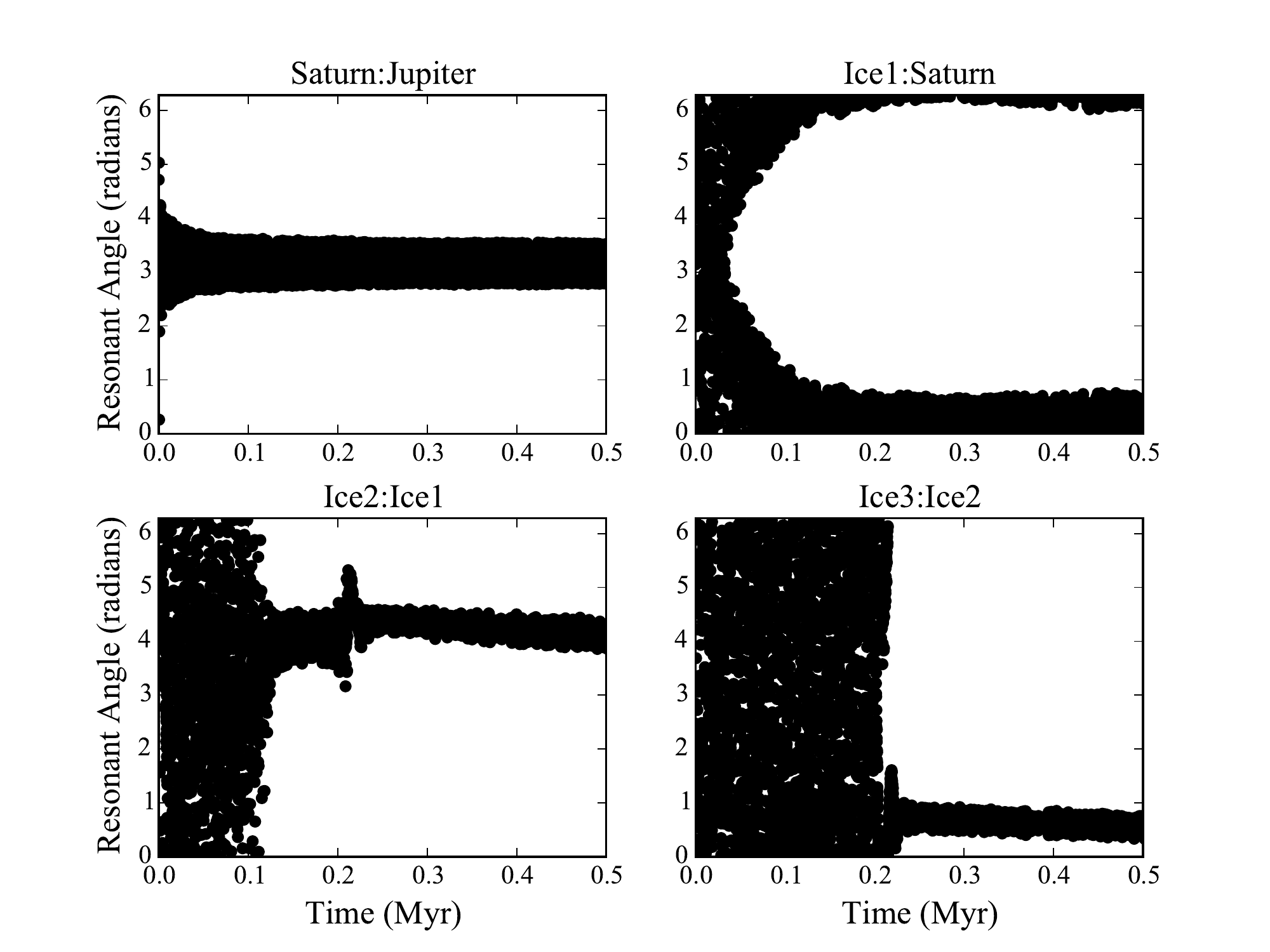}
\caption{An example of the resonant evolution of a system of outer planets in the n1 batch.  With the forcing function mimicking interactions with the gaseous disk in place, once a set of planets falls in to a MMR, they remain locked in resonance for the remainder of the evolution.}
\label{fig:res}
\end{figure}

We first evolve the gas giants (without terrestrial planets or a exterior disk of planetesimals) using the Mercury6 Bulirsch-Stoer integrator (as opposed to the faster hybrid integrator) with a 6.0 day timestep \citep{chambers99,bs}.  The Bulirish-Stoer method is necessary when building resonant configurations because the force on a particle is a function of both the positions and momenta \citep{chambers99,bat10}.  The planets are initially placed on orbits just outside their respective resonances, and then integrated until the outermost planet's semi-major axis is at the appropriate location (table \ref{table:gp}).  Figure \ref{fig:res} shows an example of this evolution for a simulation in the n1 batch.  To verify that the planets are in a MMR, libration about a series of resonant angles is checked for using the method described in \citet{clement17}.

After the resonant chain is assembled, we add a disk of 1000 equal massed planetesimals using the inner and outer radii that \citet{nesvorny12} showed best meet dynamical constraints for the outer solar system (table \ref{table:gp}).  The orbital distribution of the planetesimals is chosen in a manner consistent with previous authors \citep{bat10,nesvorny12,kaibcham16}, and follows an $r^{-1}$ surface density profile.  Angular orbital elements (arguments of pericenter, longitudes of ascending node and mean anomalies) for the planetesimals are selected randomly.  Eccentricities and inclinations are drawn from near circular and co-planar gaussian distributions (standard deviations: $\sigma_{e}=.002$ and $\sigma_{i}=.2^{\circ}$).  These same distributions are utilized throughout our study to maintain all initial eccentricities less than 0.01, and inclinations within $1^{\circ}$.  These systems are integrated using the Mercury6 hybrid integrator \citep{chambers99} with a 20.0 day timestep up until the point when two giant planets first pass within 3 mutual Hill Radii.  Because we wish to investigate specific instability times with respect to terrestrial planetary formation, and our resonant configurations can often last for tens of millions of years prior to experiencing an instability, we integrate these systems up until the onset of the instability with an empty inner solar system.  Only after this point are the terrestrial planetary formation disks at various stages of evolution embedded in these systems.  The method allows us to save computational time, and control at exactly what point the instability occurs during the giant impact phase.  Though the giant planets are already on significantly eccentric orbits at this point (and therefore affecting the terrestrial disk), we find that systems can last for millions of years before experiencing an instability if the simulation is stopped sooner.  Moreover, the terrestrial planets are far more sensitive to the evolution of Jupiter and Saturn than that of the ice giants.  We find that, in the vast majority of simulations, Jupiter and Saturn don't begin evolving substantially until after this first close encounter time.  In the vast majority  of our simulations, the instability ensues within several thousand years of the first close-encounter time.  This is consistent with simulations of planet-planet scattering designed to reproduce giant exoplanet systems \citep{chatterjee08,juric08,ray10}.

Because we want to embed terrestrial planetary disks at different stages of development into a giant planet instability, we begin by modeling terrestrial disk evolution in the presence of a static Jupiter and Saturn in a 3:2 MMR.  We form 100 systems of terrestrial planets using the Mercury6 hybrid integrator \citep{chambers99} and a 6.0 day timestep.  Because of the integrator's inability to accurately handle low pericenter passages, objects are considered to be merged with the sun at 0.1 au \citep{chambers99,chambers01}.  We choose the simplest initial orbital distributions for objects in the terrestrial forming disk in order to mirror previous studies which assumed no prior disk depletion \citep{chambers01,chambers07}.  Half of the disk mass is in 100 equal massed embryos, with the remainder in 1000 equal massed planetesimals.  The spacing of the embryos and planetesimals is selected to achieve a surface density profile that falls off radially as  $r^{-3/2}$.  Angular orbital elements are selected randomly, and eccentricities and inclinations are drawn from near circular, co-planar gaussian distributions  ($\sigma_{e}=.002$ and $\sigma_{i}=.2^{\circ}$).  The initial disk mass in simulations numbered 0-49 is set to 5 $M_{\oplus}$.  Runs numbered 50-99 begin with 3 $M_{\oplus}$ of material.  Additionally, half of the simulations begin with the inner disk edge at 0.5 au (numbers 0-24 and 50-74) as opposed to 0.7 au.  This gives us 25 simulations with each mass/edge permutation.  Additionally, the embryo spacing varies between $\sim$5-12 mutual hill radii (depending on the simulation initial conditions and disk locations), and is consistent with simulations of oligarchic growth of embryos \citep{kokubo98}.  Furthermore, terrestrial planet formation is a highly chaotic process in which the stochasticity of the actual process dominates over the effects of small changes to initial conditions such as embryo masses and spacing \citep{hoff17}.  For a summary of these initial conditions, see table \ref{table:ics}.  In all simulations the outer disk edge is set at 4.0 au.  Furthermore, we place Jupiter and Saturn in a 3:2 MMR at roughly the same orbital locations as in the n1 and n2 configurations (ice giants are not included for this portion of the simulation).  As in previous studies, we evolve each system for 200 Myr \citep{ray09a,kaibcowan15},  outputting a snapshot of each system at $10^{4}$, $10^{5}$, $10^{6}$ and $10^{7}$ years (these times roughly correspond to the time elapsed following gas disk dispersal, which we loosely correlate with the onset of the giant impact phase).  These snapshots are then input into the giant planet configurations n1 and n2 described above (therefore each output is used twice, table \ref{table:gp}), and integrated through the giant planet instability for an additional 200 Myr using the same integrator package and timestep.  

This gives us 800 different instability simulations.  We refer to the two different giant planet configurations (n1 and n2; each containing 400 individual systems) as the simulation ``set.''  We then denote each subset of 100 integrations with a unique instability delay time ($10^{4}$, $10^{5}$, $10^{6}$ and $10^{7}$ years) as ``batches.''  And finally each unique simulation within a batch (100 each) is referred to as a ``run.''  The completed terrestrial formation simulations (with a static Jupiter and Saturn in their pre-instability 3:2 MMR) become our control batch (100 runs).  Thus the control batch represents a sample of terrestrial formation outcomes in a late Nice Model scenario, where the giant planets remain locked in their mutual resonant configuration until the instability occurs $\sim$ 400 Myr after the planets finish forming.

\begin{table}
\centering
\begin{tabular}{c c c}
\hline
Run Number & Disk Mass ($M_{\oplus}$) & Disk Inner edge (au)\\
\hline
0-24 & 5.0 & 0.5 \\	
25-49 & 5.0 & 0.7 \\	
50-74 & 3.0 & 0.5 \\	
75-99 & 3.0 & 0.7 \\
\hline	
\end{tabular}
\caption{Summary of initial conditions for terrestrial planetary formation simulations.}
\label{table:ics}
\end{table}

\subsection{Instability Timing}

We relate time zero in our simulations (the beginning of our control runs) with the epoch of gas dispersal, which loosely correlates with the beginning of the giant impact phase.  However, the exact dynamical state of the terrestrial disk (for our purposes, the relative abundance of larger planet embryos within the planetesimal sea) at the time of gas disk dispersal is not exactly known.  For this reason, the specific instability times we test, and draw conclusions about, are of less importance than the particular dynamical state of the disk.  In the subsequent sections of this text, we compare broad characteristics of our early (0.01 and 0.1 Myr) and late (1 and 10 Myr) instability delay times.  Though we can confidently make a general connection between these specific instability times and the time elapsed following gas dispersal in the solar system, relating the dynamical state of the terrestrial disk with the timing of the instability is the more important conclusion of our work.  As we expand upon in the subsequent sections, the later two instability delays we test tend to be more successful than the earlier ones.  Therefore our work correlates the giant planet instability timing with a terrestrial disk at a state of evolution that is mostly depleted of small planetesimals, with most of the mass concentrated in a handful of growing planet embryos.  Subsequently overlaying this timeline on that of Mars' growth inferred from isotopic dating \citep{Dauphas11} is difficult because the relationship between gas disk dispersal and CAI (Calcium-Aluminum-rich inclusion) formation is not well known.  

Furthermore, \citet{marty17} presented evidence that cometary bombardment accounted for $\sim$ 22$\%$ of the noble gas concentration in Earth's atmosphere.  At first glance, this constraint appears to be slightly at odds with the various delay times we examine in this paper (0.01-10 Myr).  We choose to discuss this here in detail because it can potentially be construed to undermine the merit of our study.  Because the noble gas makeup of the mantle is so different from that of the atmosphere, and that of comet 67P, this seems to imply that the onslaught of comets (the timing of which would correlate with the giant planet instability) occurred after the moon forming impact.  Because the moon forming impact occurred after Mars had completed forming \citep{kleine09,Dauphas11}, the giant planet instability could not be the mass-depleting event in the Mars forming region.  However this argument does not take in to account the timing of impacts with respect to the Earth's magma ocean phase.  It is reasonable to assume that some cometary delivery must have occurred prior to core closure.  If fractionalization of Xenon occurred during the magma ocean phase, the preserved signature in the mantle could very well be different from that in the atmosphere. Because neither the distribution of impact times of primordial Kuiper belt objects (KBOs), nor the fractionalization of Xenon in the magma ocean are well known, drawing a broad conclusion of the timing of the instability from this constraint is difficult.  Additionally, delivery itself may have been stochastic; in that an early instability might set the Xenon content of the mantle by delivering many small comets, and then a later impact of a large comet could boost the Xenon fraction in the atmosphere.  Finally, Xenon isotope trends are not well known over a large enough sample of comets.  It is reasonable to expect that comet 67P's specific Xenon concentration would fall somewhere on a continuum when compared to other similar comets.  A larger sample of such measurements must be made before these conclusions can be applied to the giant planet instability timeline.  Therefore, as a starting point for our study, we argue that the most important constraint for instability timing is the survivability of the terrestrial planets, which no scenario to date can ensure.

\section{Success Criteria}

When analyzing the results of our simulations, the parameter space for comparison to the actual solar system is extensive.  Furthermore, for many metrics our accuracy is strongly limited by the resolution of our simulations.  For example, the planetary embryos used in the majority of our simulations begin the integration with a fourth of Mars' present mass.  Therefore, a ``successful'' Mars analog could be formed from as few as 2-3 impacts.  For these reasons our criteria must be broad, because we are more interested in looking at statistical consistencies and order of magnitude agreements than perfectly replicating every nuance of the actual solar system.  Thus, we focus on 10 broad criteria for replication of both the inner and outer solar systems, which we summarize in table \ref{table:crit}.

\begin{table*}
\centering
\begin{tabular}{c c c c c}
\hline
Code & Criterion &  Actual Value & Accepted Value & Justification\\
\hline
A & $a_{Mars}$ & 1.52 au & 1.3-2.0 au & Inside AB \\
A,A1 & $M_{Mars}$ & 0.107 $M_{\oplus}$ & $>0.025,<.3 M_{\oplus}$ & \citep{ray09a} \\
A,A1 & $M_{Venus}$ & 0.815 $M_{\oplus}$ & $>$0.6 $M_{\oplus}$ & Within $\sim25\%$\\
A,A1 & $M_{Earth}$ & 1.0 $M_{\oplus}$ & $>$0.6 $M_{\oplus}$ & Match Venus\\
B & $\tau_{Mars}$ & 1-10 Myr & $<$10 Myr & \\
C & $\tau_{\oplus}$ & 50-150 Myr & $>$50 Myr & \\
D & $M_{AB}$ & $\sim$ 0.0004 $M_{\oplus}$ & No embryos & \citep{chambers01} \\
E & $\nu_{6}$ & $\sim$0.09 & $<$1.0 & \\
F & $WMF_{\oplus}$ & $\sim10^{-3}$ & $>10^{-4}$ & Order of magnitude\\
G & AMD & 0.0018 & $<$0.0036 & \citep{ray09a} \\
H & $N_{GP}$ & 4 & 4 & \citep{nesvorny12} \\
I & $a_{GP}$ & 5.2/9.6/19/30au & 20$\%$ & \citep{nesvorny12} \\
I & $\bar{e}_{GP}$ & 0.046/0.054/0.044/0.01 & $<$.11 & \citep{nesvorny12} \\
I & $\bar{i}_{GP}$ & 0.37/0.90/1.02/$.67^{\circ}$ & $<2^{\circ}$ & \citep{nesvorny12} \\
J & $P_{Sat}/P_{Jup}$ & 2.49 & $<$2.8 & \citep{nesvorny12} \\

\hline
\end{tabular}
\caption{Summary of success criteria for the solar system.  The columns are: (1) the semi-major axis of Mars, (2-4) The masses of Mars, Venus and Earth, (5-6) the time for Mars and Earth to accrete $90\%$ of their mass, (7) the final mass of the asteroid belt, (8) the ratio of asteroids above to below the $\nu_{6}$ secular resonance between 2.05-2.8 au, (8) the water mass fraction of Earth, (9) the angular momentum deficit (AMD) of the inner solar system, (10) the final number of giant planets, (11-13) the semi-major axes, time-averaged eccentricities and inclination of the giant planets, (14) the orbital period ratio of Jupiter and Saturn.  A complete discussion of the success criteria, background information and justifications is provided in the Supplementary Information.}
\label{table:crit}
\end{table*}

\subsection{The Inner Solar System}
Because our goal is to look for systems like our own, with particular emphasis on forming Mars analogs, we employ an analysis metric similar to \citet{chambers01}.  A system is considered to meet criterion A if it forms a Mars sized body in the vicinity of Mars' semi-major axis, exterior to two Earth sized bodies.  We first check for any planets formed in the region of 1.3-2.0 au, where the inner edge of this region is roughly equal to Mars' current pericenter ($\sim$1.38 au) and the outer limit lies at the inner edge of the asteroid belt.  If this planet has a mass less than 0.3 $M_{\oplus}$, is immediately exterior to two planets each with masses greater than 0.6 $M_{\oplus}$, and the system contains no planets greater than 0.3 $M_{\oplus}$ in the asteroid belt, criterion A is satisfied.  A separate success criteria (criterion D, section 3.3) filters out systems that finish with an embryo in the Asteroid Belt as unsuccessful.  While some authors \citep{hansen09} select 0.2 $M_{\oplus}$ as the upper mass limit for Mars analogs, due to the previously discussed resolution limitation encountered when using 0.025 $M_{\oplus}$ embryos, we follow the prescription in \citet{ray09a} and use 0.3 $M_{\oplus}$ as our limit.  Additionally, because our simulations do not take collisional fragmentation in to account (for further discussion of this phenomenon, see section 5.6), it is possible that the masses of our Mars analogs are somewhat over-estimated.  Because we set the Venus/Earth minimum mass to 0.6 $M_{\oplus}$ (approximately 75$\%$ that of Venus' present mass), this provides an adequate mass disparity of a factor of 2 between the Venus/Earth and Mars analogs.  We also look at systems which form three planets of the correct mass (criterion A1), but do not have the correct semi major axes (eg: Mars formed at a semi-major axis greater than 2.0 au).  For this criterion, we include systems which form no Mars, but do accrete appropriately sized Earth and Venus analogs in the correct locations as being successful.  

\subsection{The Formation Timescales of Earth and Mars}
Mars is often thought to have been left behind as a ``stranded embryo'' \citep{morb00} during the process of planetary formation because the timescale for its accretion inferred from Hf/W dating (.1-10 Myr) is so quick \citep{mars,Dauphas11}.  Contrarily, Earth is believed to have formed much slower; of order 50-150 Myr \citep{earth,kleine09}.  There is a significant amount of uncertainty in both of these timescales.  The specific timing of the moon forming impact, which is thought to correlate with the last major accretion event on Earth, is still not well known.  Unfortunately, these metrics are quite difficult to meet when using standard embryo accretion numerical models.  In fact, planets with semi-major axes greater than 1.3 au in our control simulation (which assume no gas giant evolution) almost always form far $\it{slower}$ than interior planets.  With these metrics in mind, we require our Mars analogs accrete 90$\%$ of their mass within 10 Myr of the beginning of our terrestrial planetary formation simulations (not the onset of the instability, criterion B).  Additionally, we require our Earth analogs take at least 50 Myr to accrete 90$\%$ of their mass (criterion C)

\subsection{The Asteroid Belt}
Imposing strict constraints on the asteroid belt is difficult because, of the 1000 planetesimals that begin a given simulation, typically only 10 to 30 complete the integration in the asteroid belt region.  Furthermore, because the smallest objects in our simulations have masses $\sim$16 times greater than Ceres, our initial conditions are quite unrealistic for an appropriate study of the asteroid belt.  Our ability to model the depletion of the asteroid belt is thus limited.  Therefore, we cannot draw any conclusions about the total mass of the asteroid belt as all the particles in our simulation are simply too large.  However, the dynamical behavior of our small planetesimals should be roughly similar to that of the larger asteroids in the belt (such as Ceres).  Because there are only a few such large asteroids in the actual asteroid belt, it is important that we heavily deplete the region of such objects in our simulations.  Several studies have already investigated the effects of the Nice Model on the asteroid belt \citep{obrien07,walshmorb11,roig15,deienno16}.  A similar study, using tens of thousands of smaller particles in the asteroid belt region will be required to study the particular dynamical constraints our scenario places on the asteroid belt.  Furthermore, the most successful models for the asteroid belt \citep{walsh11,ray17sci} successfully reproduce the compositional dichotomy between ``S-types'' (Silicate rich, moderate albedo asteroids) and ``C-types'' (low albedo, carbonaceous asteroids making up about $75\%$ of the belt) \citep{gradie82}.  Improving our mass resolution within the asteroid belt in the future will allow us to test this constraint as well.

For our purposes, we simply require that no embryos remain in the region (a $>$ 2.0 au).  This method (criterion D) is similar to that employed in \citet{chambers01} and \citet{ray09a}.  The fact that there are no significant gaps between observed mean motion and secular resonances in the actual asteroid belt implies an upper limit (about a Mercury mass) for the mass of the largest object in the belt that could survive terrestrial planetary formation \citep{obrien11}.  Because the total mass of the asteroid belt is thought to have depleted by about a factor of $\sim10^{4}$ over the life of the solar system \citep{petit01}, a detailed calculation of the actual numerical value of what is left over is far beyond the scope of this paper.  Moreover, because Gyr timescale modeling of test particles in the asteroid belt indicates that depletion is logarithmic over the life of the solar system, the majority of this depletion must happen during the first $\sim$200 Myr of evolution because loss in the next 4 Gyr is only of order $\sim$50$\%$ \citep{minton10}.

Additionally, we look at the number of remaining planetesimals above and below the $\nu_{6}$ secular resonance between 2.05-2.8 au.  In the actual solar system, this ratio is about $\sim.09$.  Again, due to the small number statistics, the ratio inferred from an individual simulation will be very imprecise.  Therefore we only require this ratio to be less than one (criterion E).

Furthermore, resonance sweeping during giant planet migration and evolution can drastically effect the dynamical structure of the asteroid belt \citep{walshmorb11,minton11,roig15}.  In a slow migration scenario, as Saturn moves outward towards its current semi-major axis, the $\nu_{16}$ secular resonance excites inclinations as it sweeps through the asteroid belt.  As Saturn continues to migrate, the $\nu_{6}$ resonance erodes the remaining low inclination, low eccentricity component.  Though the process of resonance sweeping can undoubtedly have an effect on the resulting mass of Mars \citep{bromley17}, the mechanism can also remove low inclination asteroids which are common (figure \ref{fig:4}) in today's asteroid belt \citep{walshmorb11}.  To preserve the structure of the asteroid belt from the effects of resonance dragging, previous authors \citep{roig15} utilized a ``jumping Jupiter'' model instability (wherein Jupiter and Saturn ``jump'' toward their present orbital locations, ideally preserving the fragile terrestrial planets and asteroid belt structure).  In order for our model to be successful, it must heavily deplete the asteroid belt while still maintaining the low inclination component.

\subsection{Water Delivery to Earth}
Many models trace the origin of Earth's water to the early depletion of the primordial asteroid belt \citep{ray07,ray09a}.  The actual topic of water delivery to Earth is extensive, with many competing models, and far beyond the scope of this paper (for a more complete discussion of various ideas see \citet{morb00}, \citet{morb12} and \citet{marty16}).  Uncertainties in the initial disk properties and locations of various snow lines make it challenging for embryo accretion models like our own to confidently quantify the water mass fraction (WMF) of Earth analogs.  In addition, the actual bulk water content of Earth is extremely uncertain.  Estimates of the mantle's water content range between 0 to tens of oceans \citep{lecuryer98,marty12,halliday13}, awhile the core may contain 0 to nearly 100 \citep{badro14,nomura14}.  See the review by \citet{hirschmann06} for a discussion of the difference between the capacity of Earth's water reservoirs and geochemical evidence for the actual water contained in Earth's interior.  Furthermore, given the amount of planetesimal scattering which occurs when the giant planets grew and migrated during the gas disk phase, the material from which Earth formed during the giant impact phase may have already been sufficiently water rich \citep{ray17}. For our simulations, we first look at the bulk WMF of Earth analogs calculated using an initial water radial distribution similar to that used in \citet{ray09a} (equation \ref{eqn:wmf}, this assumes that the primordial asteroid belt region was populated by water-rich objects from the outer solar system during the gas disk phase \citep{ray17}).  Any system which boosts Earth's WMF to greater than $10^{-4}$ is considered to satisfy criterion F.  We also analyze the percentage of objects Earth analogs accrete from different sections of the disk.

\begin{equation}
WMF=
\begin{cases}
10^{-5},\quad r<2au\\
10^{-3},\quad 2au<r<2.5au\\
10\%,\qquad r>2.5au
\end{cases}
\label{eqn:wmf}
\end{equation}

\subsection{Angular Momentum Deficit}
One defining aspect about our solar system is the remarkably low eccentricities and inclinations of the terrestrial planets.  Over lengthy integrations, the orbits of all the inner planets but Mercury typically stay extremely low \citep{quinn91,laskar09}.  This orbital constraint was very difficult for early accretion models to meet \citep{chambers98,chambers01}.  \citet{obrien06} used dynamical friction to explain how the orbits could stay cold through the standard process of planetary formation.  However, it has proven even more challenging to keep eccentricities low when a giant planet instability is considered \citep{bras09,kaibcham16}.  Any successful model of terrestrial planetary formation must maintain low orbital excitation in the inner solar system.  To measure this in our systems, we measure the angular momentum deficit (AMD, criterion F) of each system \citep{laskar97}.  AMD (equation \ref{eqn:amd}) quantifies the deviation of the orbits in a system from perfectly coplanar, circular orbits.  We follow the same procedure as \citet{ray09a} and require our systems maintain an AMD less than twice the value of the modern inner solar system ($\sim$.0018).

\begin{equation}
	AMD = \frac{\sum_{i}m_{i}\sqrt{a_{i}}[1 - \sqrt{(1 - e_{i}^2)}\cos{i_{i}}]} {\sum_{i}m_{i}\sqrt{a_{i}}} 
	\label{eqn:amd}
\end{equation}

\subsection{The Outer Solar System}
When analyzing the success of our terrestrial planetary formation simulations, it is important to consider how dependent our results are on the fate of the outer solar system.  Indeed, the chance of our chosen resonant chains reproducing all the important traits of the outer solar system after undergoing an instability is often low.  For example, \citet{nesvorny12} report only a 33$\%$ chance of our n1 configurations finishing the integration with the correct number of outer planets.  When all four success criteria in that work are considered, n1 resonant chains only successfully match the outer solar system $\sim4\%$ of the time.  Given the computational cost of our integrations, we are less interested in how $\it{often}$ we correctly replicate the orbital architecture of the giant planets, and more concerned with how $\it{dependent}$ our results are on the fate of the outer solar system.

To quantify the outer solar system, we adopt the same success criteria as \citet{nesvorny12}.  First, criterion H requires that the simulation finish with 4 giant planets.  If this is satisfied, criterion I stipulates that the final semi-major axis of each planet be within 20$\%$ of the modern location, and the time averaged eccentricity and inclination of each planet be less than twice the largest current value in the outer solar system.  Finally, criterion J states that the period ratio of Jupiter and Saturn stay less than 2.8.  It should be noted that, unlike \citet{nesvorny12}, we check for criterion J independently of whether the other two standards are met.  The dynamics of the forming terrestrial planets are far less sensitive to the behavior of the ice giants than they are to that of Jupiter and Saturn.  Therefore, we are nearly just as interested in systems that correctly produce Jupiter and Saturn but eject too many ice giants as we are in those that replicate the outer solar system perfectly.

\section{Results and Discussion}

\begin{table*}
\centering
\begin{tabular}{c c c c c c c c c}
\hline
Set & A & A1 & B & C & D & E & F & G\\ 
& $a,m_{TP}$ & $m_{TP}$ & $\tau_{mars}$ & $\tau_{\oplus}$ & $M_{AB}$ & $\nu_{6}$ & WMF & AMD\\
\hline
Control & 0 & 0 & 9 & 86 & 2 & 53 & 87 & 8\\
n1/.01Myr & 3 & 15 & 31 & 84 & 41 & 35 & 40 & 14\\
n2/.01Myr & 2 & 6 & 15 & 75 & 48 & 46 & 34 & 9\\
n1/.1Myr & 0 & 2 & 16 & 79 & 38 & 39 & 27 & 12\\
n2/.1Myr & 6 & 10 & 20 & 69 & 50 & 39 & 38 & 12\\
n1/1Myr & 13 & 13 & 6 & 90 & 38 & 31 & 47 & 13\\
n2/1Myr & 8 & 14 & 11 & 87 & 54 & 42 & 44 & 7\\
n1/10Myr & 12 & 20 & 3 & 73 & 48 & 26 & 47 & 16\\
n2/10Myr & 8 & 20 & 19 & 80 & 54 & 31 & 62 & 8\\
\hline
\end{tabular}
\caption{Summary of percentages of systems which meet the various terrestrial planet success criteria established in table \ref{table:crit}.  It should be noted that, because runs beginning with a disk mass of 3$M_{\oplus}$ were not successful at producing appropriately massed Earth and Venus analogs, criterion A and A1 are only calculated for 5$M_{\oplus}$ systems.  The subscripts TP and AB indicate the terrestrial planets and asteroid belt respectively.}
\label{table:results}
\end{table*}

We provide complete summaries of our results in tables \ref{table:results}, \ref{table:results_GP} and \ref{table:results_2.8}.  It should be noted that a small number of our instabilities fail to properly eject an ice giant, and complete the integration with greater than four giant planets.  Because we are only interested in systems similar to the solar system, we do not include these few outliers in any of our analyses.  Table \ref{table:results} shows the total percentage of systems in each simulation batch which meet our success criteria for the inner solar system.  Table \ref{table:results_GP} summarizes the percentage of systems satisfying our giant planet success criteria.  We find that our systems adequately replicate the outer solar system with frequencies consistent with those reported in \citet{nesvorny12}.  In table \ref{table:results_2.8}, we look at our success rates for systems with Jupiter/Saturn configurations most similar to the actual solar system (period ratios less than 2.8).  Clearly, the fate of the terrestrial planets is highly dependent on the evolution of the solar system's two giant planets.  This is largely due to strong secular perturbations which result from the post-instability excitation of the giant planet orbits.  Indeed, when we look at systems which eject all ice giants and finish with a highly eccentric Jupiter and Saturn outside a period ratio of 2.8, we find terrestrial planets which are too few in number, on excited orbits and systematically under-massed.  Though we still see these symptoms in some of the systems summarized in table \ref{table:results_2.8}, they are noticeably less frequent.

\begin{table}
\centering
\begin{tabular}{c c c c}
\hline
Set & H & I & J \\ 
& $N_{GP}$ & a,e,$i_{GP}$ & S:J \\
\hline
n1/.01Myr & 27 & 14 & 47\\
n2/.01Myr & 18 & 6 & 41\\
n1/.1Myr & 14 & 7 & 42\\
n2/.1Myr & 18 & 3 & 38\\
n1/1Myr & 23 & 15 & 57\\
n2/1Myr & 22 & 9 & 38\\
n1/10Myr & 17 & 9 & 36\\
n2/10Myr & 16 & 8 & 36\\

\hline
\end{tabular}
\caption{Summary of percentages of systems which meet the various giant planet success criteria established in table \ref{table:crit}.  The subscript GP indicates the giant planets.}
\label{table:results_GP}
\end{table}

\begin{table*}
\centering
\begin{tabular}{c c c c c c c c c}
\hline
Set & A & A1 & B & C & D & E & F & G\\ 
& $a,m_{TP}$ & $m_{TP}$ & $\tau_{mars}$ & $\tau_{\oplus}$ & $M_{AB}$ & $\nu_{6}$ & WMF & AMD\\
\hline
Control & 0 & 0 & 9 & 86 & 2 & 53 & 87 & 8\\
n1/.01Myr & 0 & 15 & 33 & 94 & 24 & 33 & 61 & 15\\
n2/.01Myr & 5 & 5 & 17 & 80 & 30 & 60 & 56 & 10\\
n1/.1Myr & 0 & 0 & 14 & 78 & 13 & 36 & 42 & 5\\
n2/.1Myr & 12 & 18 & 7 & 84 & 32 & 59 & 64 & 11\\
n1/1Myr & 26 & 26 & 12 & 95 & 20 & 29 & 65 & 14\\
n2/1Myr & 11 & 27 & 12 & 92 & 42 & 44 & 69 & 2\\
n1/10Myr & 9 & 18 & 7 & 92 & 16 & 29 & 53 & 25\\
n2/10Myr & 20 & 33 & 27 & 87 & 25 & 52 & 77 & 2\\
\hline
\end{tabular}
\caption{Summary of percentages of systems which meet the various terrestrial planet success criteria established in table \ref{table:crit} $\bf{\it{AND}}$ finish with Jupiter and Saturn's period ratio less than 2.8 (criterion J).  It should be noted that, because runs beginning with a disk mass of 3$M_{\oplus}$ were not successful at producing appropriately massed Earth and Venus analogs, criterion A and A1 are only calculated for 5$M_{\oplus}$ systems.  The subscripts TP and AB indicate the terrestrial planets and asteroid belt respectively.}
\label{table:results_2.8}
\end{table*}

\subsection{Formation of a Small Mars}

An early instability is highly successful at producing a small Mars, regardless of instability timing and the particular evolution of the giant planets.  $75\%$ of all our instability systems form either no Mars or a small Mars (less than 0.3 $M_{\oplus}$), as opposed to none of our control runs.  Additionally, as shown in table \ref{table:results}, most of our control systems leave at least one embryo in the asteroid belt.  In fact, many of these systems even form multiple small planets, or an Earth massed planet in the asteroid belt.  Only $9\%$ of our instability simulations form a planet more massive than Mars in the asteroid belt, as opposed to $65\%$ of our control runs.  Clearly a Nice Model instability is a highly efficient means of depleting the planetesimal disk region of material outside of 1.3 au.

\begin{figure}
\centering
\includegraphics[width=.5\textwidth]{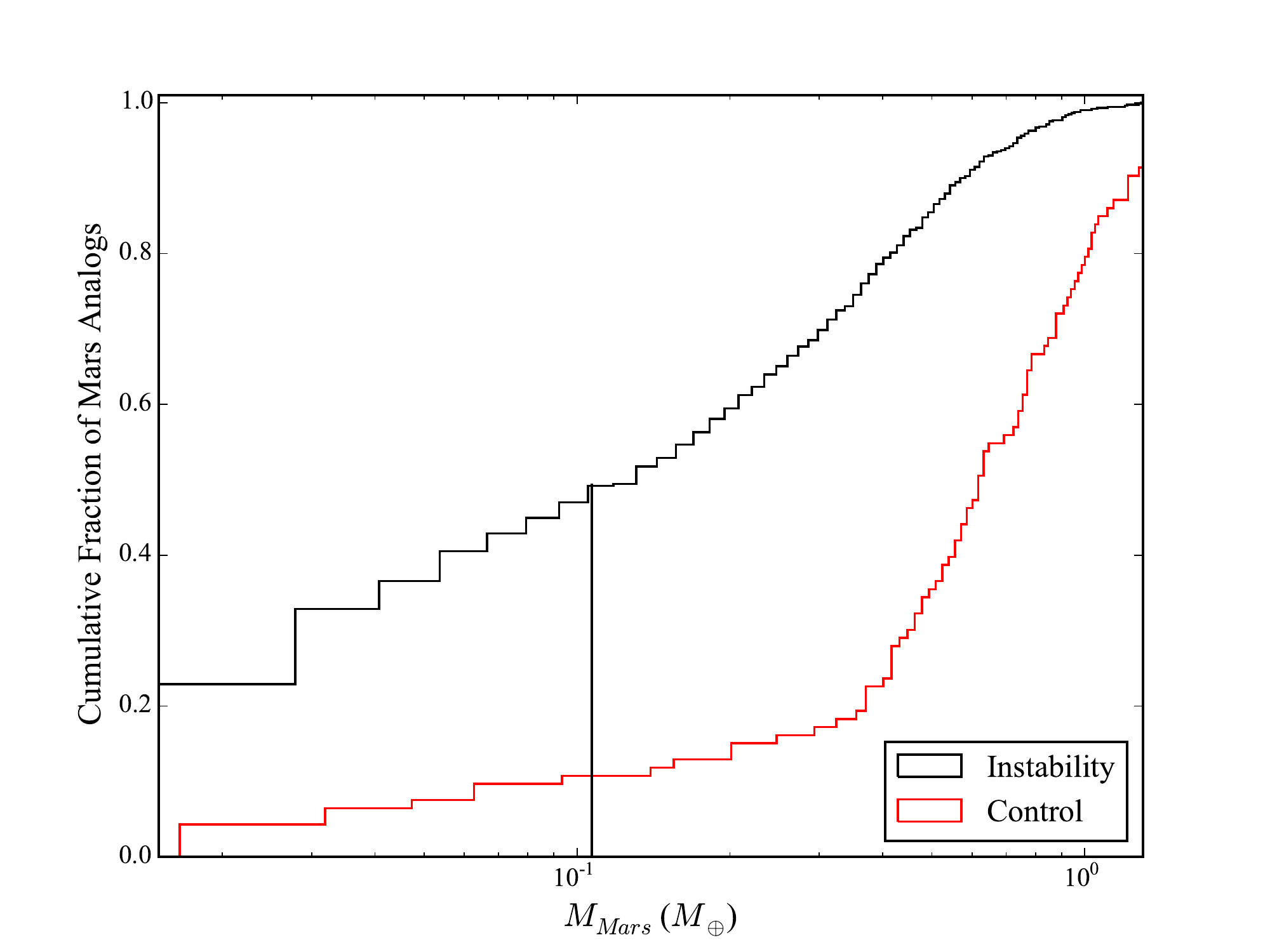}
\caption{Cumulative distribution of Mars analog masses formed in instability systems and our control batch (note that some systems form multiple planets in this region, here we only plot the largest planet).  The vertical line corresponds to Mars' actual mass.  All control runs with a Mars analog smaller than 0.3 $M_{\oplus}$ ($\sim$20$\%$ of the batch) were unsuccessful in that they also formed a large planet in the asteroid belt}
\label{fig:mars}
\end{figure}

\begin{figure}
\centering
\includegraphics[width=.5\textwidth]{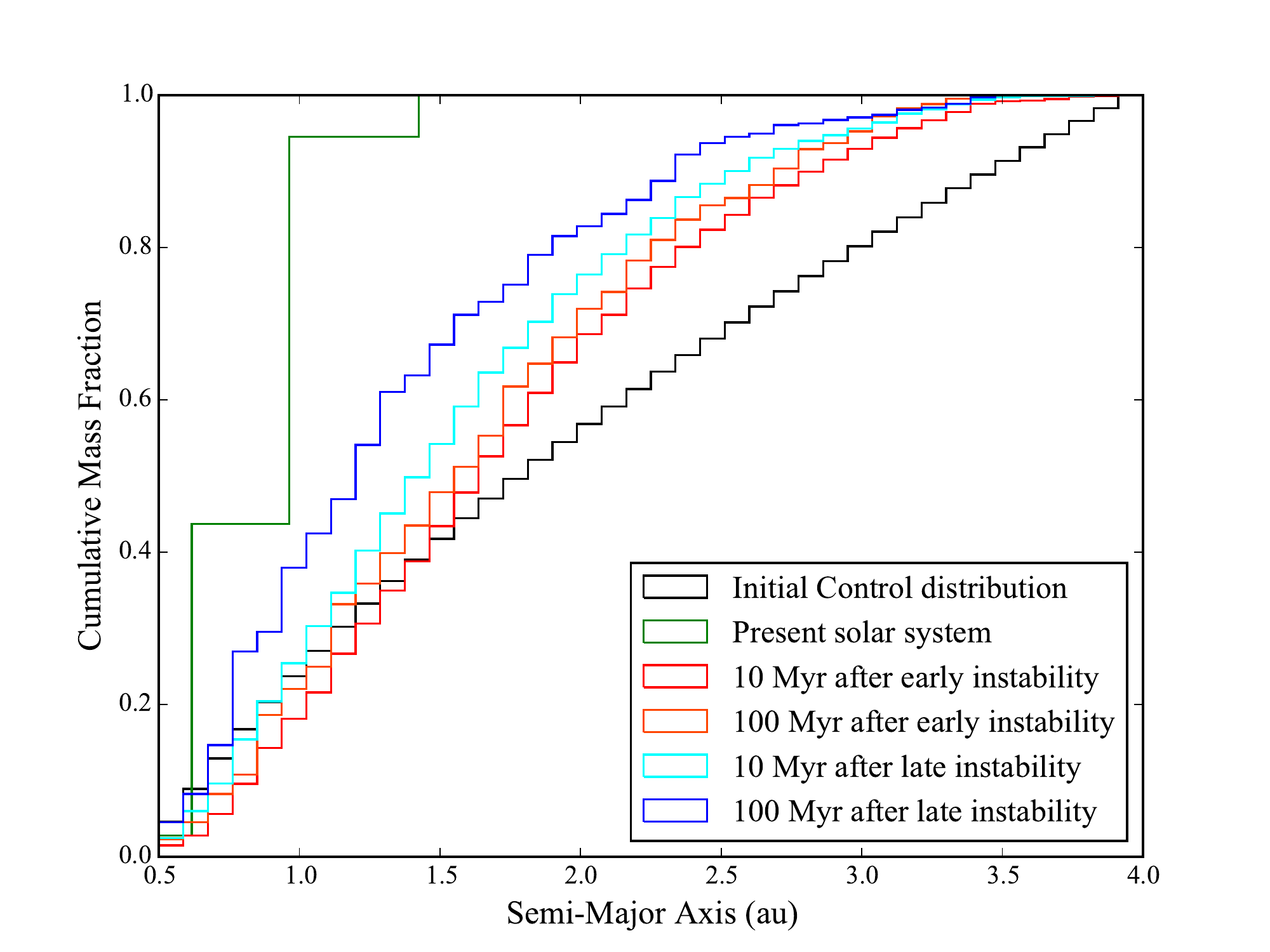}
\caption{Cumulative mass distribution of embryos and planetesimals at the beginning of the control simulations (black line), 10 Myr after the early instability delay times (.01 and 0.1 Myr; red line), 100 Myr after the early instability delay times (orange line), 10 Myr after the late instability delay times (1 and 10 Myr; cyan line) and 100 Myr after the late instability delay times (blue line).  The green line represents the current mass distribution of the inner solar system.}
\label{fig:mass}	
\end{figure}

Figure \ref{fig:mars} shows the cumulative distribution of the largest planets in each system formed between 1.3 and 2.0 au for our instability sets versus the control batch.  The solar system fits in well with this distribution, with slightly greater than half of our systems forming Mars analogs larger than the actual planet.  Indeed, the instability consistently starves this region of material and produces a small planet.  In fact, 22$\%$ of our systems produce no planet in the Mars region whatsoever.  This is slightly lower for the two earliest instability delays (0.01 and 0.1 Myr) which we test, with 18$\%$ of such systems forming no Mars in the 1.3 to 2.0 au region.  This is due to the fact that, when the instability occurs, the ratio of the number of planetesimals to embryos, and that of total planetesimal mass to total embryo mass is much higher.  In the late instability cases, the majority of the system mass is trapped in several large embryos.  The dynamical excitement of the additional planetesimals in the early instability delay cases allows the disk mass to disperse, thereby enhancing the mass of Mars analogs.  We find that planets in the outer disk (a $>$ 1.3 au) from early instability delay systems (0.01 and 0.1 Myr) where the outer planets meet criterion J accrete $\sim6$ times more material from the inner disk than in late instabilities (1 and 10 Myr).  Furthermore, dynamical friction from the higher number of planetesimals de-excites material in the outer disk.  Indeed, our early instability delay systems which satisfy criterion J lose an average of 0.25 $M_{\oplus}$ less mass in the outer disk (a $>$ 1.3 au) to ejection or collisions with the Sun than the later delays.  Figure \ref{fig:mass} shows how a late instability delay results in a dramatically steeper mass distribution profile. 

\begin{figure*}
\centering
\includegraphics[width=.49\textwidth]{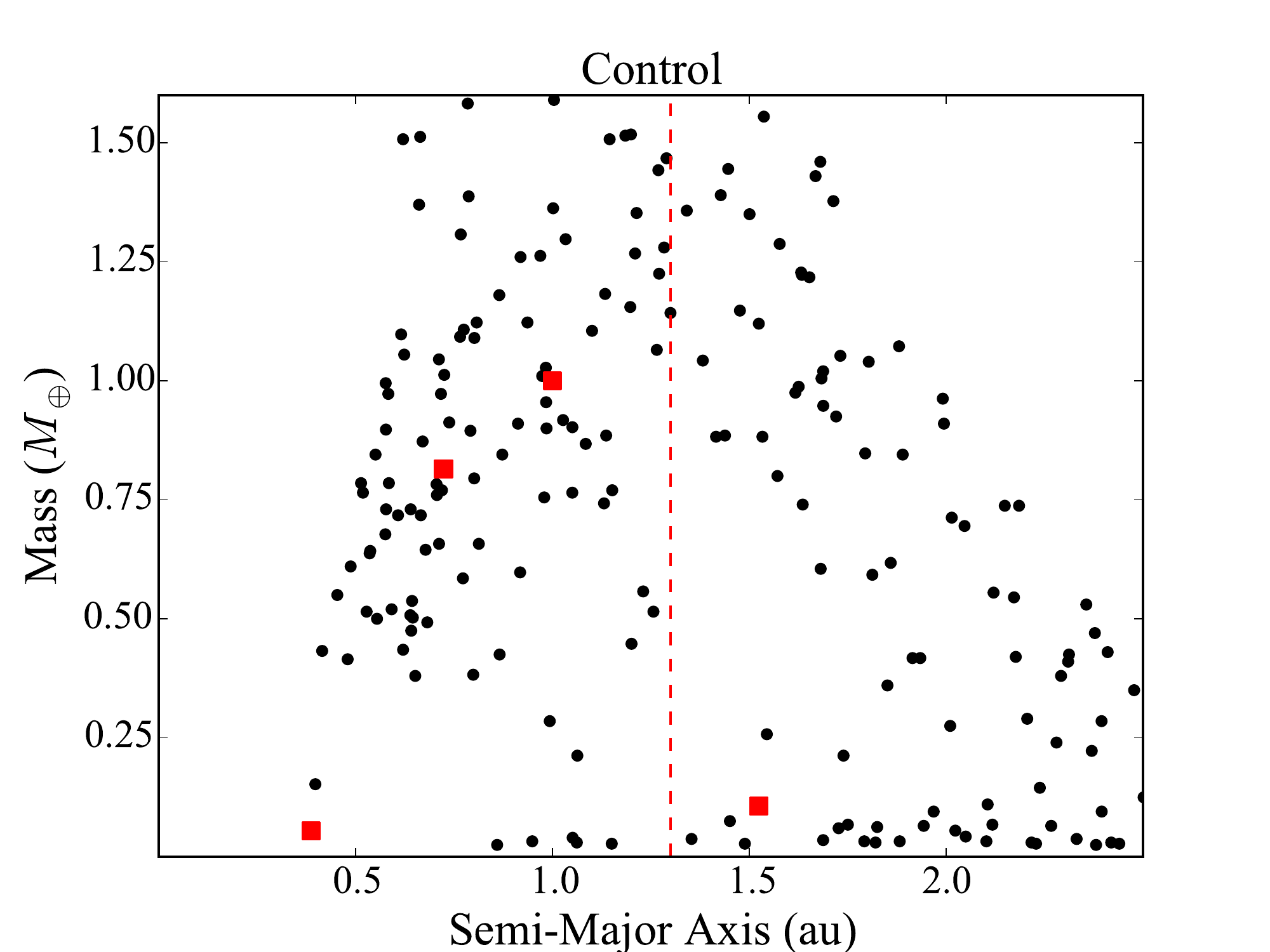}	
\qquad
\includegraphics[width=.49\textwidth]{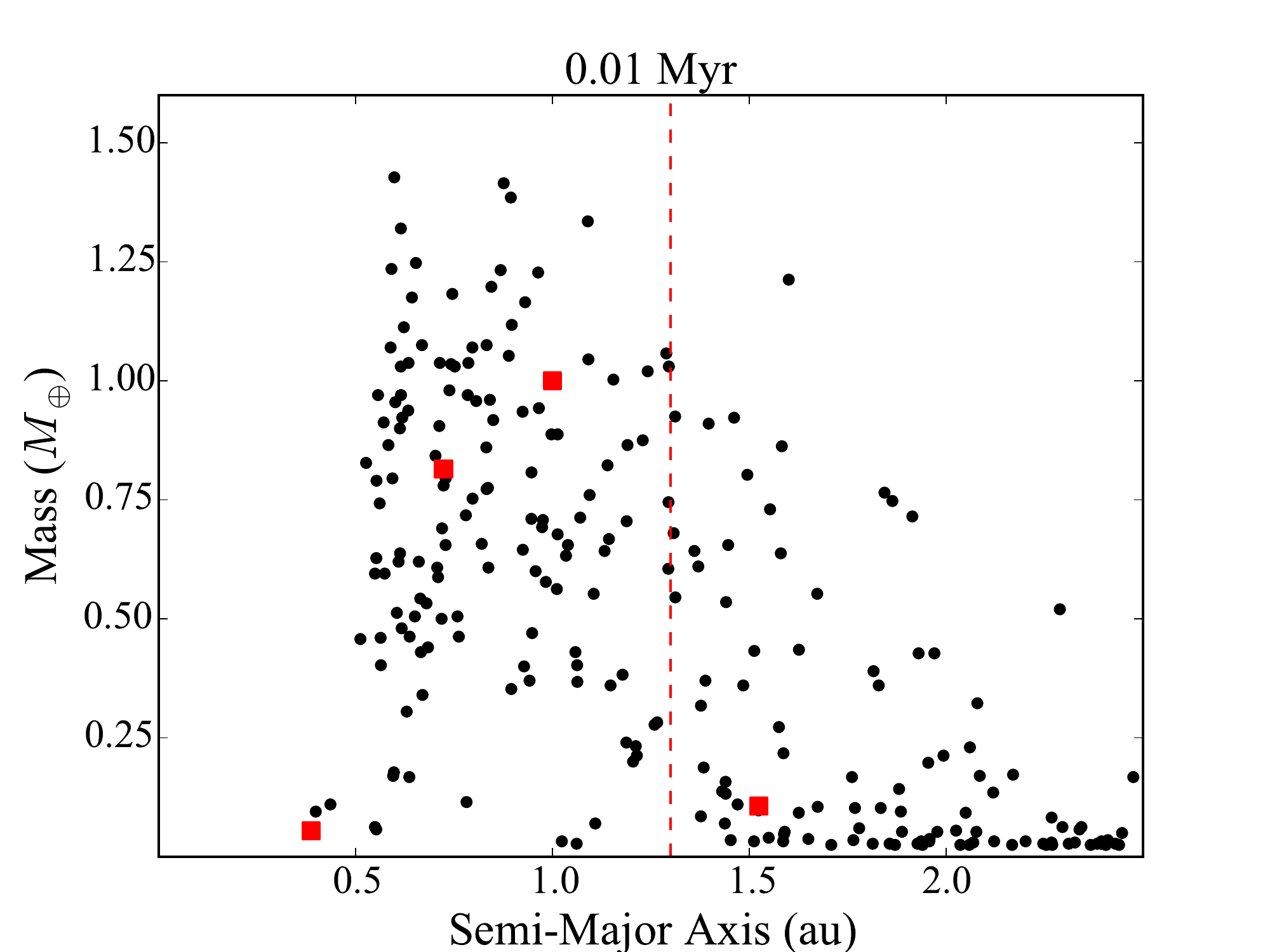}	
\includegraphics[width=.49\textwidth]{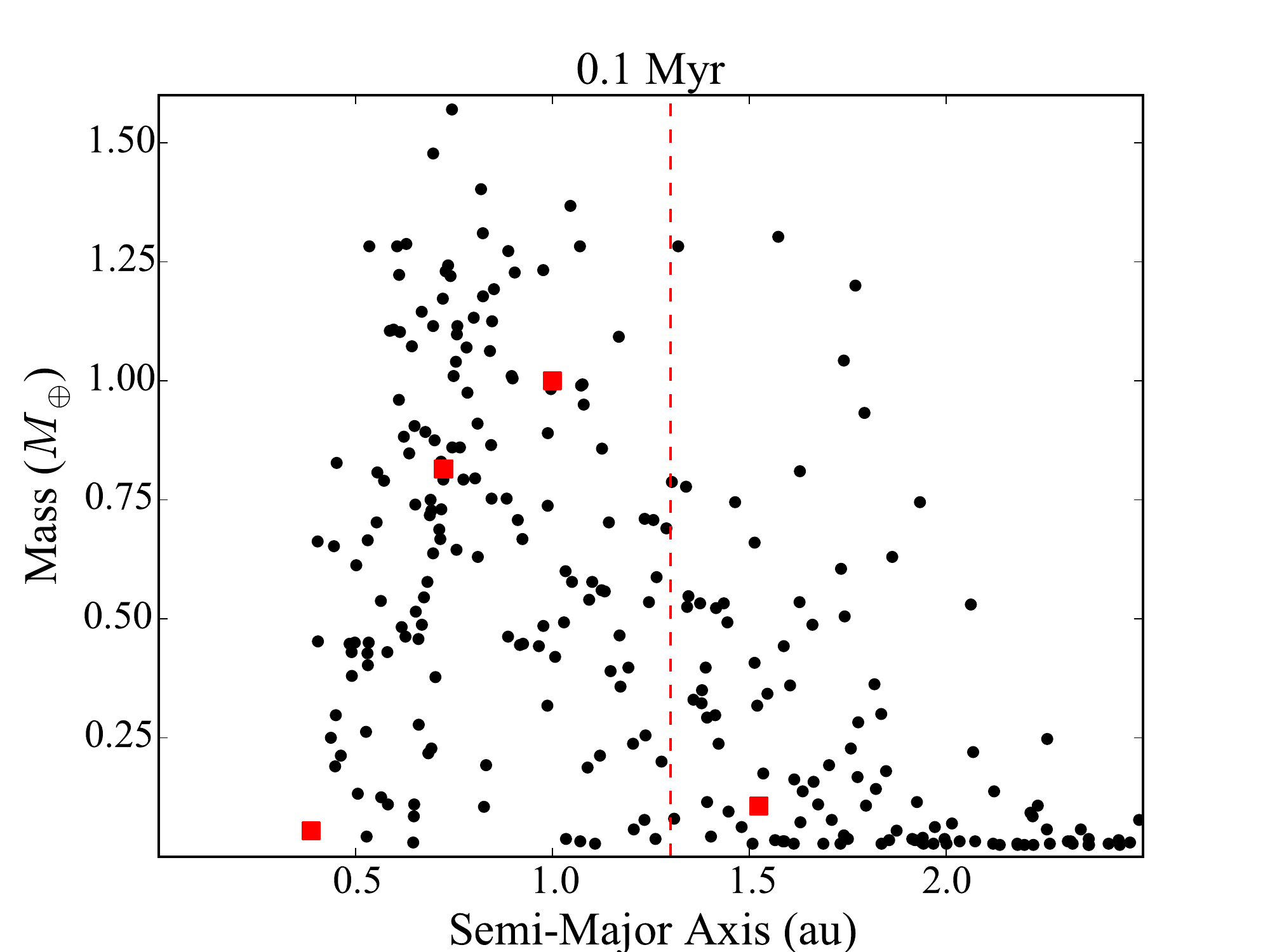}	
\qquad
\includegraphics[width=.49\textwidth]{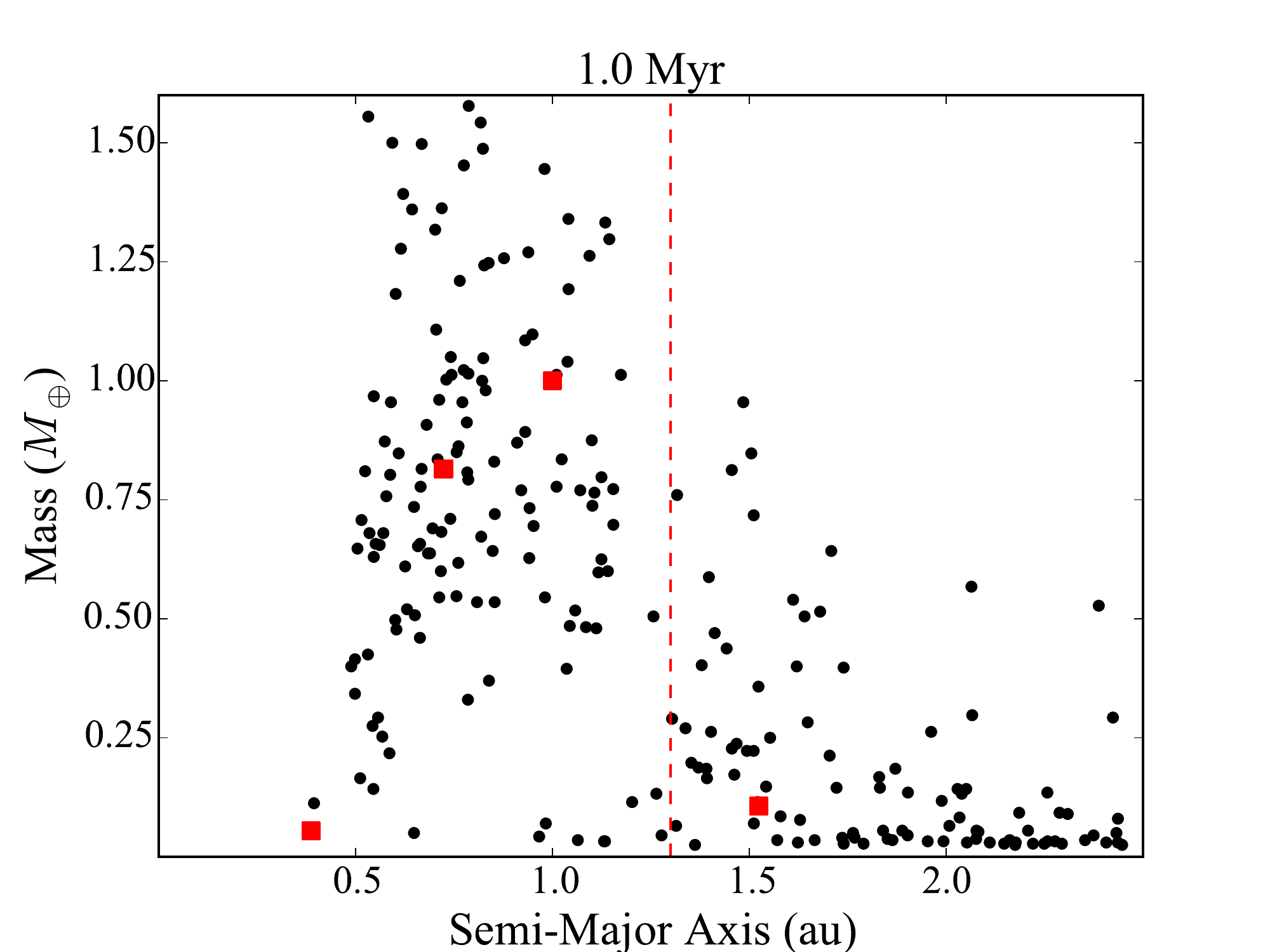}	
\includegraphics[width=.49\textwidth]{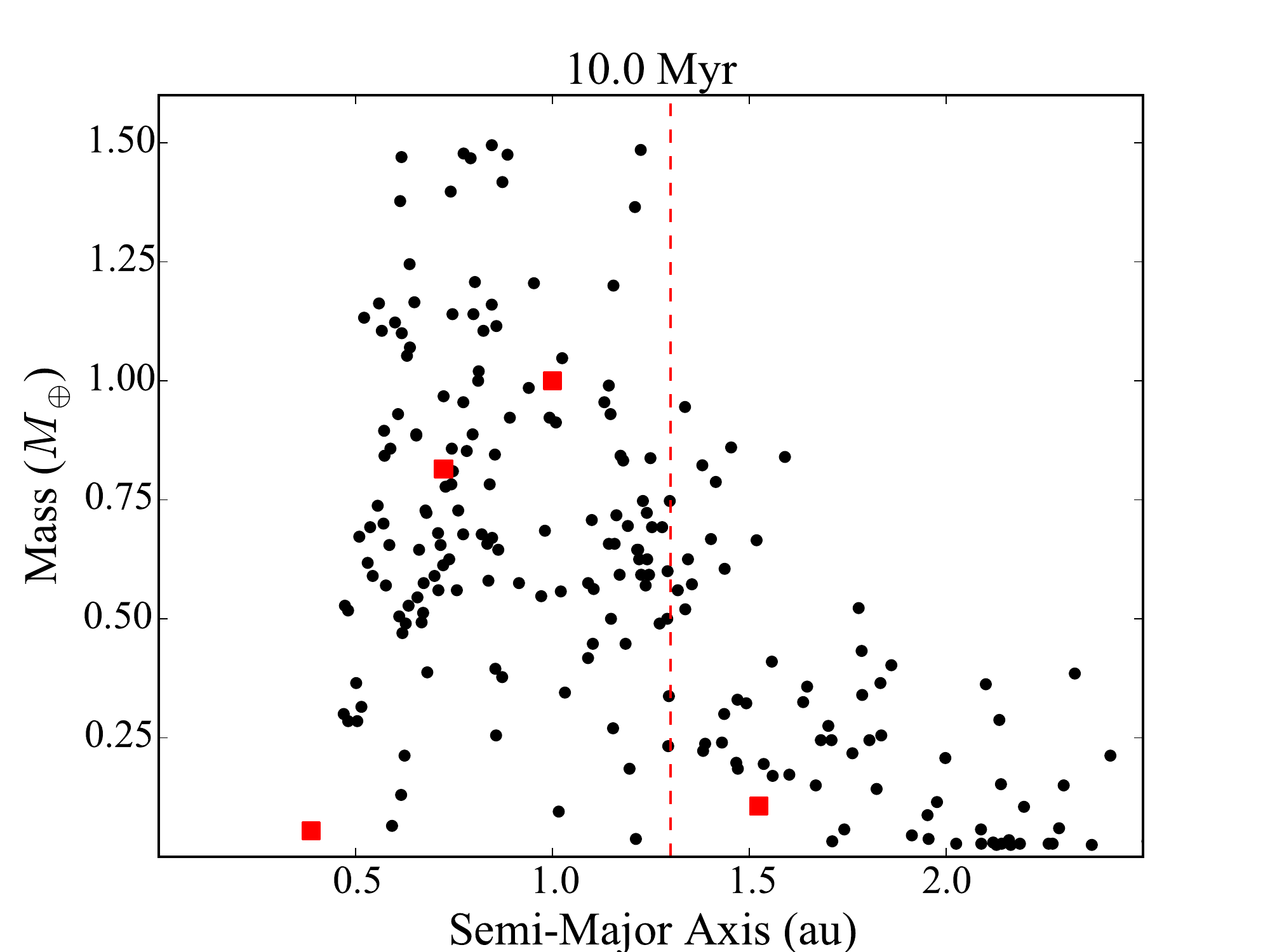}	
\caption{Distribution of semi-major axes and masses for all planets formed using 5 $M_{\oplus}$ massed planetesimal disks.  The red squares denote the actual solar system values for Mercury, Venus, Earth and Mars.  The vertical dashed line separates the Earth and Venus analogs (left side of the line) and the Mars analogs (right side).  The top panel shows our control runs and each of the 4 lower plots depict a different instability delay time.}
\label{fig:totalplot}
\end{figure*}

In Figure \ref{fig:totalplot} we show the distributions of semi-major axes and masses for the planets we form, compared with our control simulations.  Regardless of the instability delay time, there is a stark contrast between our simulations and the control set.  Earth mass planets in the Mars region and beyond are very common in the control simulations, and rarely occur when the system undergoes a Nice Model instability.  Though the general trends for all four plots are quite similar, we note that our distributions for late instabilities are slightly better matches to the actual solar system for two reasons.  First, the number of outlying Mars analogs which are larger than $\sim.6 M_{\oplus}$ is substantially less for the later instability delay times.  As discussed previously, because these simulations begin with fewer planetesimals, the lack of disk dispersal and de-excitation via dynamical friction between small bodies makes it difficult for the system to accrete a large Mars over the next $\sim$ 200 Myrs of evolution.  Next, these systems tend to form more accurate Earth and Venus analogs.  Many of the failed early instability delay simulations are clear examples of the instability's tendency to hinder the formation of Earth and Venus.  The larger dispersal of the disk mass profile in these delays consistently deprives the Earth and Venus forming regions of material.  In these simulations, we often form systems with small (less than $\sim.6 M_{\oplus}$) Venus and Earth analogs, just one total terrestrial planet, or no inner planets at all.

\subsection{Strengths of an Early Instability}

Tables \ref{table:results} and \ref{table:results_2.8} show the percentage of systems in each batch that meet our success criteria for the inner solar system.  Because the solar system very well could have been formed in a low likelihood scenario, it is important not to place too much weight on meeting specific numerical values and exactly replicating every particular dynamical trait of the actual system.  For this reason, we try to keep our success criteria as broad as possible.

We consider our systems roughly successful at meeting our established criteria.  Generally, our systems perform better than our control runs in almost all categories.  Because the unique dynamical state of the solar system represents just one point in a broad spectrum of possible outcomes, it is unreasonable to expect that our simulations meet every single success criterion exactly, every time.  By these standards, our simulations are successful on most accounts.  In particular, an instability is very successful at meeting the requirements for the asteroid belt (criterion D) and the formation timescale of Earth (criterion C).  

\begin{figure}
\includegraphics[width=.5\textwidth]{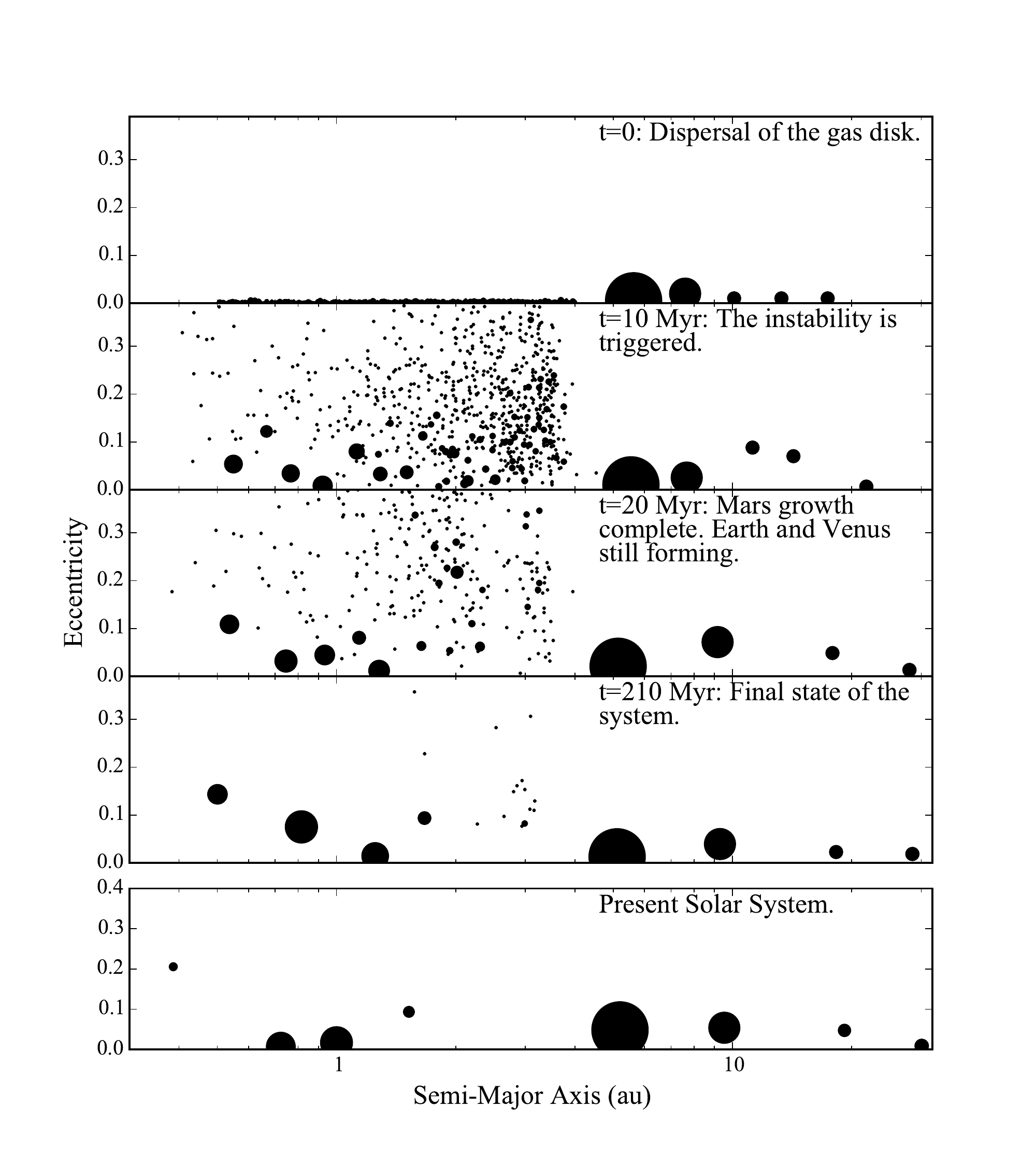}
\caption{Semi-Major Axis/Eccentricity plot depicting the evolution of a successful system in the n1/10Myr batch.  The size of each point corresponding to the mass of the particle (because Jupiter and Saturn are hundreds of times more massive than the terrestrial planets, we use separate mass scales for the inner and outer planets).  The final planet masses are 0.37, 1.0, 0.69 and 0.15 $M_{\oplus}$ respectively.}
\label{fig:2}
\end{figure}

Given the large number of constraints involved in criterion A, our success rates of $\sim5-20\%$ are still very encouraging.  For this reason, we also use the broader criterion A1 for the orbital architecture of the inner solar system.  This metric considers systems that form planets of the correct mass ratios, but incorrect orbital locations, and those which form no Mars but a proper Earth and Venus pair to be successful.  When we look at our rates of success for meeting this criterion when Jupiter and Saturn finish the integration within a period ratio of 2.8 (Table \ref{table:results_2.8}), we find our later instability delay times are remarkably successful with values closer to $\sim30\%$.  In these successful scenarios, Mars often forms as a stranded embryo.  The simulation begins with multiple bodies of order 0.25-2.5 $M_{Mars}$ in the vicinity of Mars' present orbit.  When the instability ensues, most of these bodies are ejected.  40$\%$ of the time, ``Mars'' undergoes no further major accretion events with other embryos after the instability simulation begins.  Figure \ref{fig:2} shows an example of such an evolution scheme.  Notice that after the instability ensues the proto-Venus and proto-Earth continue to accrete material while objects in the Mars forming region do not.

\subsubsection{The Asteroid Belt}

Our simulations are successful at depleting the asteroid belt because of the dynamical excitation provided by the embryos we place in the belt.  The embryos pre-excite the asteroid belt during the evolution leading up to the instability.  When the instability ensues, excited planetesimals in the belt scatter off the embryos, leading to high mass loss.  To test this, we performed a follow-on suite of integrations using our 1 Myr instability control disks, and the Mercury6 hybrid integrator.  We place Jupiter and Saturn on orbits corresponding to a period ratio of 1.6, and set an extra ice giant immediately exterior to Saturn.  When the ice giant scatters and is ejected, Jupiter and Saturn jump.  Next, systems where the post-jump period ratio of Jupiter and Saturn is between 2.1 and 2.4 are selected, and integrated for an additional 10 Myr with a code that mimics smooth migration and eccentricity damping on $\sim$ 3 Myr e-folding timescales using fictitious forces \citep{lee02}. To attain final states similar to Jupiter and Saturn, we shut off migration and eccentricity damping when the two gas giants attain a period ratio above 2.45 and eccentricities below 0.06.  Through this process, we create a sample of asteroid belts ($\sim$ 20) which experience a pre and post-instability evolution broadly similar to the runs from our original simulations that best matched the currently observed orbital architecture of Jupiter and Saturn.  By performing 2 sets of runs (embryos and planetesimals and planetesimals only), we are able to test the effects of embryo excitation.  Planetesimal only simulations are created by converting all embryos with a $>$ 1.5 au in a given system in to an appropriate number of equal-mass planetesimals with similar semi-major axes, eccentricities and inclinations, and random angular orbital elements.  Simulations using embryos and planetesimals lost about twice as much mass beyond 1.5 au over just 10 Myr of evolution as the planetesimal only systems.  This is consistent with the idea that the dynamical excitement of embryos leads to significantly more mass loss in the asteroid belt.

\begin{figure}
\includegraphics[width=.5\textwidth]{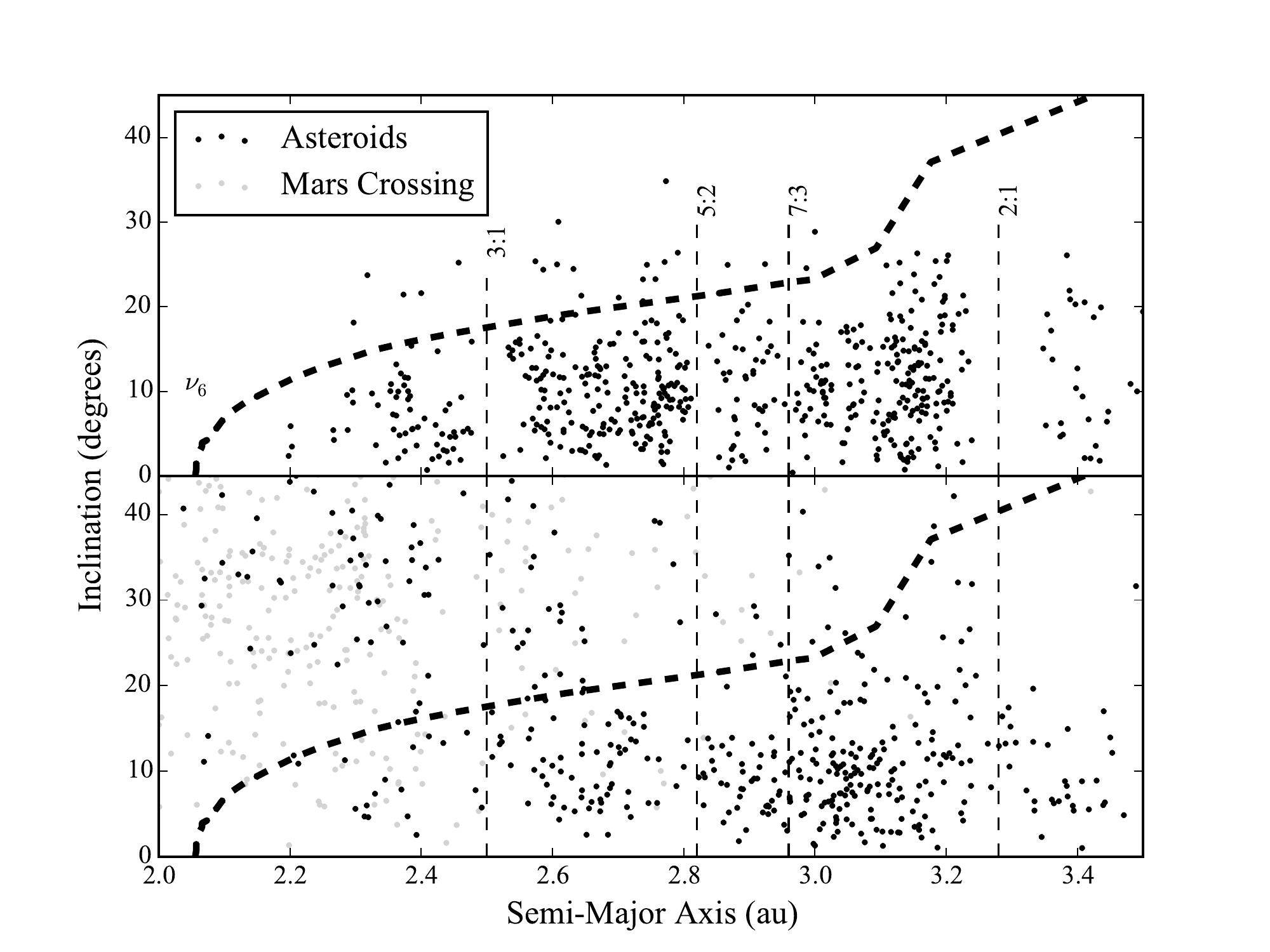}
\caption{The upper plot shows the inclination distribution of the modern asteroid belt (only bright objects with absolute magnitude H $<$ 9.7, approximately corresponding to D $>$ 50 km, are plotted). The bottom plot combines all planetesimals remaining in the asteroid belt region from all instability simulations that form a Mars analog less massive than 3 times Mars' actual mass, and finish with Jupiter and Saturn's period ratio less than 2.8.  Grey points correspond to high-eccentricity asteroids on Mars crossing orbits which will be naturally removed during subsequent evolution up to the solar system's present epoch.  The vertical dashed lines represent the locations of the important mean motion resonances with Jupiter.  The bold dashed lines indicate the current location of the $\nu_{6}$ secular resonance.}
\label{fig:4}
\end{figure}

Our simulations' ability to replicate the asteroid belt population about the $\nu_{6}$ resonance is subject to numerical limitations.  Our simulations start with 0.0025 and 0.0015 $M_{\oplus}$ planetesimals, both of which are more massive than the entire present contents of the asteroid belt.  Furthermore, most simulations finish with between 10 and 30 bodies in the main belt.  Such small numbers makes it difficult to discern subtle dynamical features within our individual simulated asteroid belts. Although statistics can be improved by co-adding many simulations to examine the general effects of giant planet instabilities (Figure \ref{fig:4}), every instability is unique at some level, and dynamical sculpting processes occurring during some instabilities may not operate in others.  When we co-add the asteroids from all of our criterion J satisfying simulations (those which finish with Jupiter and Saturn's period ratio less than 2.8) and remove objects on planet-crossing orbits, we find a ratio of bodies above to below the $\nu_{6}$ resonance (between 2.05 and 2.8 au) to be $\sim$ 0.71.  However, when we only consider asteroids between 2.05 and 2.5 au, the ratio is a poorer match (2.24).  Though neither number is close to the actual ratio ($\sim$ 0.09), the first is quite promising with respect to other numerical modeling attempts.  For example, \citet{deienno16} imposed a ``Grand-Tack'' style migration on the asteroid belt and found a ratio of $\sim$ 1.2. In a similar manner, \citet{walshmorb11} reported a ratio of $\sim$ 5.2 in a smooth migration scenario.

Due to numerical limitations, further simulations, involving tens of thousands of smaller bodies in the asteroid belt region are required to comprehensively study the detailed effect of an early instability on the asteroid belt.  However, an early instability seems to generate an asteroid belt similar to the actual belt in broad strokes.  The presence of embryos appears to provide sufficient dynamical excitation to substantially deplete the mass in the region (most simulations deplete more than 95$\%$ of belt material in 200 Myr).  Though this does fall short of the required depletion of a factor of $\sim10^{4}$ \citep{petit01}, our mechanism does produce substantial depletion in the asteroid belt when compared with our control runs.  More realistic initial conditions and handling of collisions will be required to more accurately model depletion in the Asteroid Belt in our model.  Nevertheless, embryos remaining in the asteroid belt are extremely rare in our instability systems.  By using a full instability, rather than a smooth migration scenario, we avoid dragging resonances across the belt, thus broadly preserving its orbital structure.

\subsection{Weaknesses of an Early Instability}

On average, our instability simulations are less successful at meeting the success criteria for the formation timescale of Mars, the WMF of Earth and the AMD of the terrestrial planets.  Reproducing the formation timescale of Mars is a difficult constraint for N-body accretion models of terrestrial planetary formation.  A successful Mars analog in our simulations need only be composed of 4 embryos.  Meanwhile, the real Mars formed from millions of smaller objects that accreted prior to and during the giant impact phase.  This difference must be weighed when considering moderate discrepancies between the formation timescales of simulated Mars analogs and the real planet.  In fact, $\sim$ 40$\%$ of all our Mars analogs undergo no impacts with other embryos following the instability, and Mars' form on average $\sim39$ Myr faster than their Earth counterparts.  Additionally, 6 of the 7 Mars analogs in the criterion A1 satisfying [n2/10 Myr] batch (our most successful simulations), form in under 10 Myr.  Because Mars' growth only continued at the $\sim$ 10$\%$ level after $\sim$ 2-4 Myr \citep{Dauphas11}, our 1 Myr instability delays are the most successful at simultaneously matching the mass distribution of the terrestrial system and the proposed accretion history of Mars.  However, the geological accretion history of Mars is inferred relative to CAI formation; the timing relative to gas disk dissipation of which is not fully understood. 

Providing a means of water delivery to Earth is not a strict requirement for the success of an embryo accretion model.  It should be noted that many ideas for how Earth was populated with water exist, several of which have nothing to do with delivery via bodies from the outer solar system \citep{morb00,morb12}.  In fact, water delivering planetesimals may have been scattered on to Earth-crossing orbits during the giant planets' growth and migration phase \citep{ray17}.  Despite the small number statistics involved with using only 1000 initial particles in the Kuiper belt, about half of which typically deplete in the initial phase of integration (before the terrestrial disks are imbedded), we do find 12 instances of Earth analogs accreting objects from this region in our simulations.  Interestingly, Earth's noble gases are thought to come primarily from comets, despite the fact that comets are likely a minor source of water \citep{marty16}.  We find that a late instability delay time (1 and 10 Myr) systematically stretches the feeding zone of Earth analogs further in to the terrestrial disk.  This broader feeding zone is basically a result of eccentricity excitation of planetesimals \citep{levison03}.  In these cases, mass from the outer disk is able to ``leap-frog'' its way towards the proto-Earth.  In the first phase of evolution (before the instability), forming embryos in the middle part of the disk ($\sim$2.0-3.0 au) accrete material from the outermost section of the disk ($\sim$3.0-4.0 au).  When the instability ensues, these embryos are destabilized, and occasionally scattered inward towards the forming ``Earth''.  

Our simulations often leave the inner planets with too large of an AMD.  Though most runs only exceed the actual AMD of the solar system by a factor of 2-3, some systems occasionally reach AMDs as high as 10 times the value of the current solar system.  Many of these outliers are from integrations where particularity violent instabilities leave behind a system of overly excited giant planets.  Even when we remove these instances which are not analogous to the actual solar system, our ``successful'' simulations still tend to possess high AMD values.  Often, an overly excited Mars is the source of this orbital excitation (values of $e_{Mars}\sim.1-.25$ are typical for these systems).  The obvious source of this excitation is secular interactions with the excited giant planets.  One potential solution to this problem might be accounting for collisional fragmentation.  \citet{chambers13} showed that angular momentum exchange resulting from hit-and-run collisions noticeably reduces the eccentricity of planets formed in embryo accretion models.  Additionally, \citet{jacobson14} showed that the AMD of systems increases as the total amount of initial mass placed in embryos instead of planetesimals increases, and as individual embryo mass decreases.  We observe a similar relationship in overall disk mass loss (section 5.1).  Moreover, because of the chaotic nature of the actual solar system, its AMD can evolve by as much as a factor of 2 in either direction over Gyr timescales \citep{laskar97}.

\subsection{Varied Initial Conditions}

Our simulations are broken up into 4 different sets of 25 runs with unique inner disk edge and initial disk mass combinations (Table \ref{table:ics}).  In half of our simulations, we use a disk mass of 3 $M_{\oplus}$ rather than a more typical choice of $\sim5 M_{\oplus}$ \citep{chambers01,ray09a}.  100$\%$ of these systems with lower mass disks fail to meet criterion A for correctly replicating the semi-major axes and masses of the terrestrial planets.  Using a lower overall disk mass leads to less dynamical friction available to save bodies from loss after the instability.  We find that by far the most likely final configuration for these simulations is a single Venus analog, occasionally accompanied by a Mars analog.  However, we note that the percentages of systems that meet the other 6 success criteria (criterion B through G) are roughly similar (within $\sim5\%$) for systems of either initial disk mass.

We see no noticeable differences between the sets of simulations which truncate the inner planetesimal disk at 0.5 au and those with an inner edge at 0.7 au.  Both batches are roughly equally likely ($9\%$ and $10\%$ of the time, respectively) to form a Mercury analog (we define this as any planet smaller than 0.2 $M_{\oplus}$ interior to an Earth and a Venus analog).  For more discussion on the formation of Mercury, see section 6.  Finally, our rates for meeting all success criteria for the inner planets are roughly the same (within $\sim5\%$), regardless of the selected inner disk edge location.

\subsection{Excitation of Jupiter's $g_{5}$ Mode.}

\begin{figure}
\includegraphics[width=.5\textwidth]{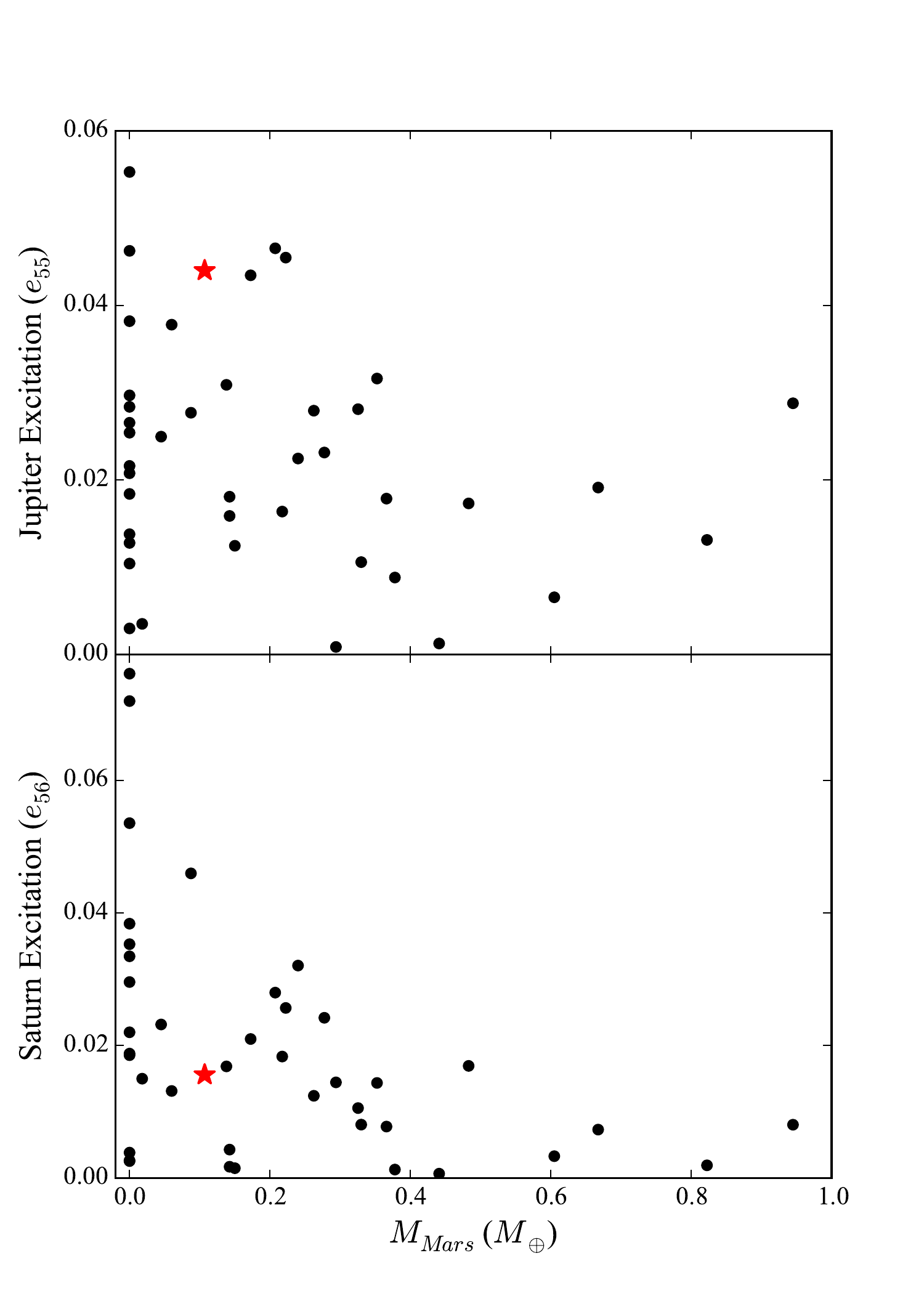}
\caption{Values of the amplitudes of Jupiter's $g_{5}$ mode versus the mass of Mars analogs formed for 10 Myr delayed instability systems where the orbital period ratio of Saturn to Jupiter completing the integration less than 2.8.  The red stars correspond with the present solar system values.}
\label{fig:3}
\end{figure}

\begin{figure}
\centering
\includegraphics[width=.5\textwidth]{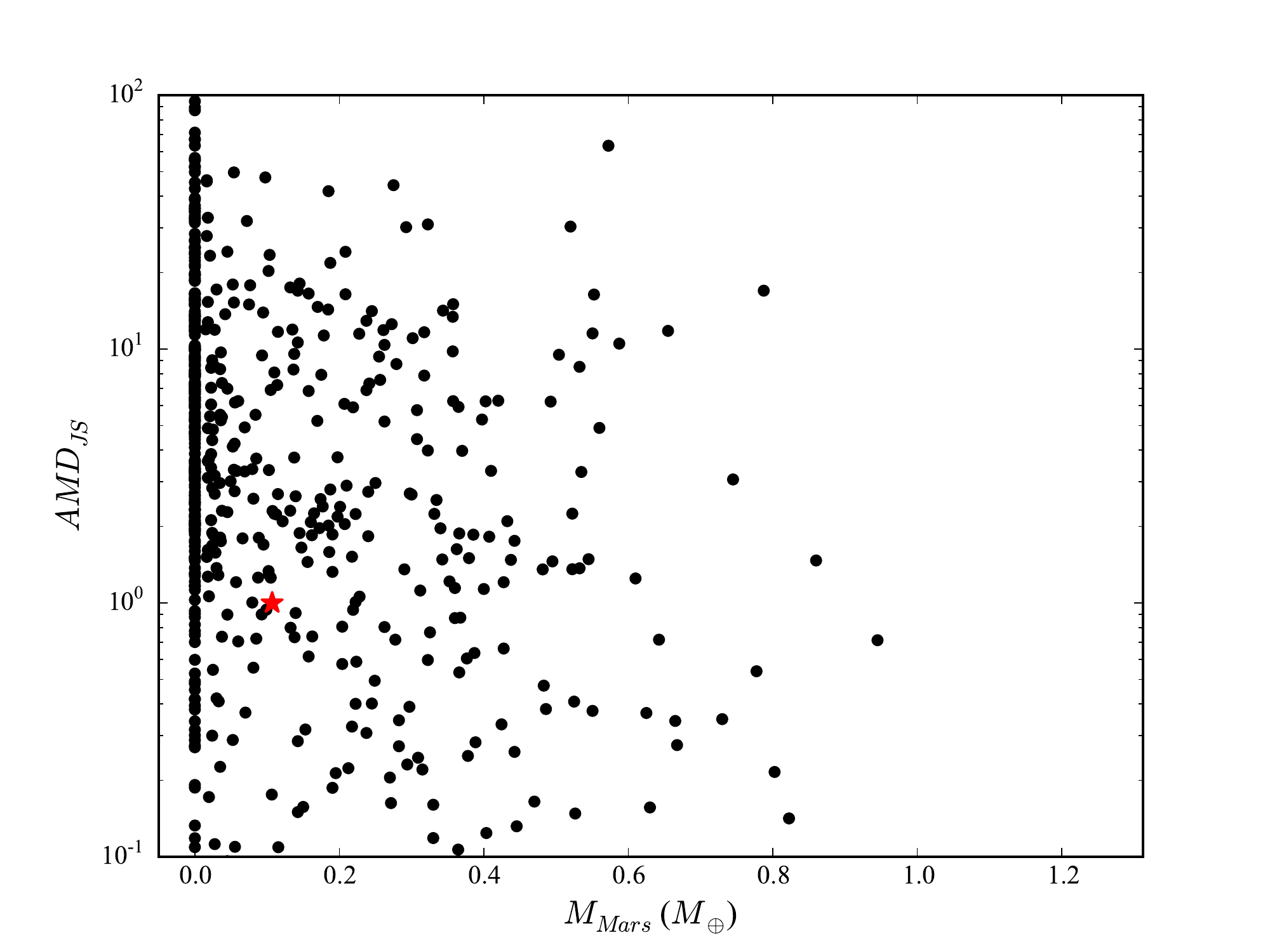}
\caption{The AMD of Jupiter and Saturn (normalized to the actual solar system value) versus the mass of Mars analogs formed for all instability systems.  The red star denotes the solar system values.}
\label{fig:amd}	
\end{figure}

\begin{figure}
\centering
\includegraphics[width=.5\textwidth]{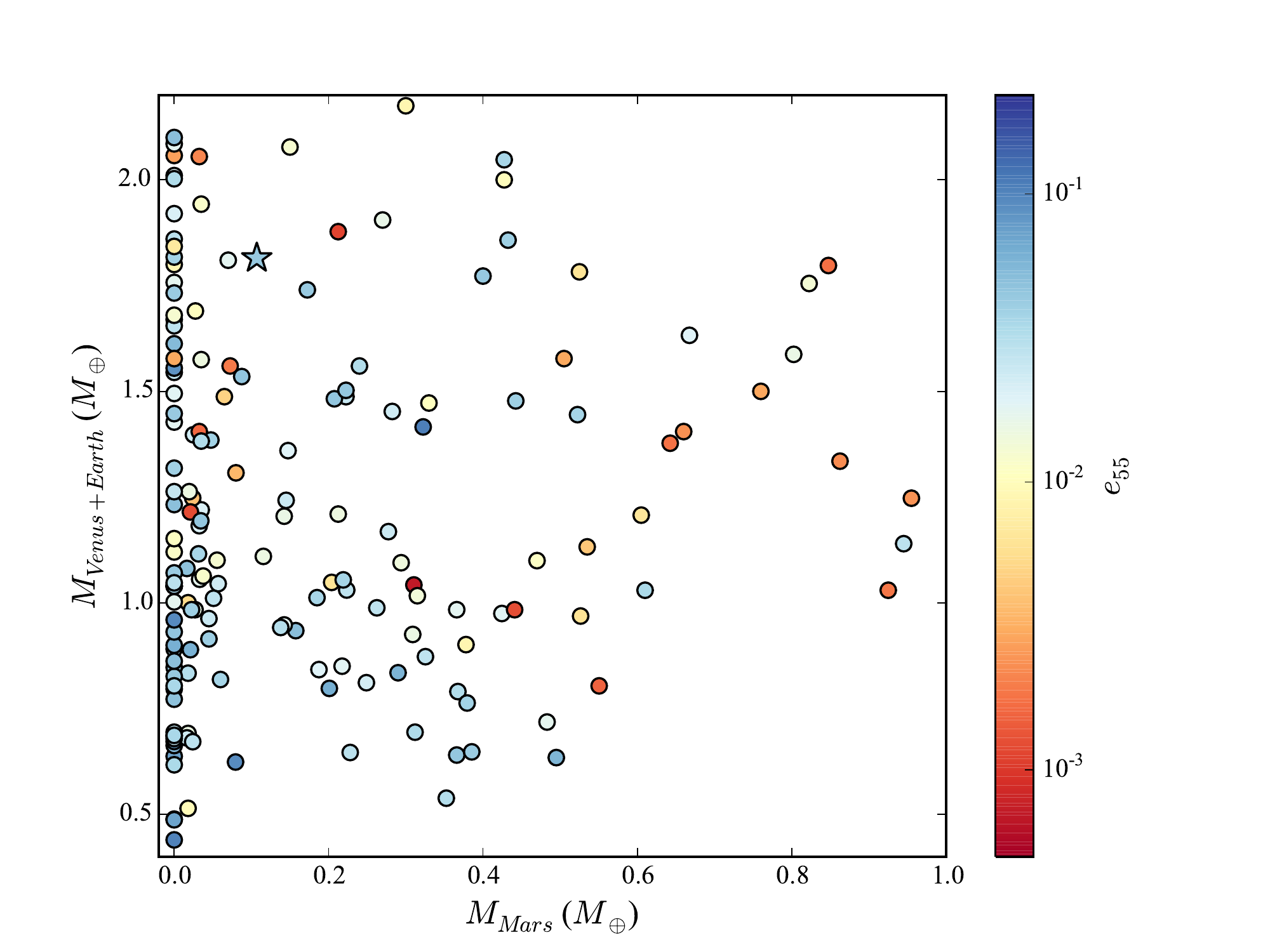}
\caption{Values of the total mass of Earth and Venus analogs versus the mass of Mars analogs formed in instability systems where the value of Jupiter and Saturn's period ratio finished the simulation less than 2.8.  The color of each point corresponds to the amplitude of Jupiter's $g_{5}$ mode.  The blue star denotes actual solar system values.}
\label{fig:sweet}	
\end{figure}

The sufficient excitation of Jupiter's $g_{5}$ mode is another important constraint on the evolution of the giant planets.  The current amplitude of the mode, $e_{55} = 0.044$, is very important in driving the secular evolution of the solar system \citep{morb09}.  Additionally, the amplitude of Saturn's forcing on Jupiter's eccentricity, $e_{56} = 0.016$, is important for the long term evolution of Mars and the asteroid belt.  Overexciting $e_{56}$ might lead to a small Mars and a depleted asteroid belt.  However, this scenario is not akin to the actual evolution of the solar system.  To evaluate the relationship between the $g_{5}$ mode and the mass of Mars, we integrate all systems which finish with Jupiter and Saturn within a period ratio of 2.8 for an additional 10 Myr, and perform a Fourier analysis of the additional evolution \citep{nesvorny96}.  In Figure \ref{fig:3}, we plot the values of $e_{55}$ and $e_{56}$ against the masses of Mars analogs produced for these systems in our 10 Myr delayed instabilities.  We find that systems with an $e_{55}$ amplitude greater than that of actual solar system never produce a large Mars analog (greater than 0.3 $M_{\oplus}$).  The average Mars analog mass in systems with $e_{55}$ less than half the solar system value (0.022) is 1.96 times Mars' mass, compared to 1.03 for systems with $e_{55} > 0.022$.  Additionally, we see multiple examples of systems where  $e_{56}$ is close to the solar system value, that produce a small Mars.  Clearly, the complete excitation of the $g_{5}$ mode is linked to reducing the mass of planets in the Mars forming region.  Additionally, in Figure \ref{fig:amd}, we plot the normalized AMD of Jupiter and Saturn versus the mass of Mars analogs.  It is very apparent that the range of possible values is extensive, with the solar system falling well within the range of our results.  Therefore, the actual solar system is consistent with our dynamical evolution model.  Furthermore, Jupiter's excitation also effects Earth and Venus.  When we plot the cumulative mass of Earth and Venus against the mass of Mars (Figure \ref{fig:sweet}), we find that many of the systems with similar values to the solar system have correspondingly similar values of $e_{55}$.

\subsection{Impact Velocities}

Our simulations use an integration scheme where all collisions are assumed to be perfectly accretionary \citep{chambers99}.  This provides a decent approximation of the final outcome of terrestrial planet formation for low relative-velocity collisions between objects with a large mass disparity.  However, higher velocity collisions can often be erosive \citep{genda12}, particularly when the projectile to target mass ratio is closer to unity.  Additionally, depending on the parameters of the impact, glancing blows can lead to the re-accretion of either all, some or none of the original projectile \citep{asphaug06,asphaug10,lands12}.  Because of the instability's tendency to excite small planetesimals on to high-eccentricity orbits, the collisional velocities in our simulations are often quite large (occasionally in excess of 10 times the mutual escape velocity).  Because of this, it is very important to consider the effects of collisional fragmentation of bodies when analyzing our results.

To check our simulations for erosive collisions, we use a code which determines the collision type from the collision speed and impact angle by following the parameter space of gravity dominated impacts mapped by \citet{lands12}.  We find that erosive collisions do occur with the forming planets in our simulations.  However, they are infrequent, and comprise less than $\sim5\%$ of all collisions and less than $\sim1\%$ by mass.  Erosive collisions occur at similar rates for Earth, Venus and Mars analogs, and are almost always planetesimal-on-embryo impacts.  We do note, however, that our Mars analogs undergo a significantly higher number of hit and run collisions (36$\%$ by mass as opposed to less than 20$\%$ for Earth and Venus analogs).  This indicates that our Mars analog masses are most likely over-estimated.  Additionally, because the effects of hit and run collisions have been shown to reduce the AMD of planets produced \citep{chambers13}, it is possible that the resulting orbital eccentricities and inclinations of our Mars analogs are similarly over-excited.  This is encouraging because the excitation of Mars significantly contributes to our systematically high AMDs.

\section{Conclusions}

In this paper, we have presented 800 direct numerical simulations of a giant planet instability occurring in conjunction with the process of terrestrial planet formation.  By timing this violent event within the first $\sim$100 Myr following the dispersion of the gas disk, the instability scenario no longer requires a mechanism to prevent the destabilization and loss of the fully formed terrestrial planets terrestrial planets (such as the ``Jumping Jupiter'' model).  When we scrutinize our fully formed systems against a wide range of success criteria, we note multiple statistical consistencies between our simulated planets and the actual terrestrial system.  First, our Mars analogs are more likely to form small and quickly.  In fact, 75$\%$ of all our instabilities form either no planet in the Mars region whatsoever, or an appropriately sized Mars.  Additionally, cases where the instability is delayed 1-10 Myr after the beginning of the giant impact phase tend to be more successful than earlier timings ($<$1 Myr).  In many of these runs, the instability itself sets the geological formation timescale of Mars.  Thus, an early giant planet instability provides a natural explanation for how Mars survived the process of planet formation as a ``stranded embryo.''

We find that our simulated asteroid belts are largely depleted of mass when compared with our control set of simulations, and seldom form a planet in the belt region.  Furthermore, the broad orbital distribution of the asteroid belt seems to be well matched when we co-add the remaining asteroids from all of our simulations.  Because the instability itself is inherently chaotic, each resulting system of giant planets has slightly different orbital characteristics.  When we filter out systems where the giant planets orbits most closely resemble those in the actual solar system, we find higher rates of success among the corresponding terrestrial systems.   

At first glance, certain geochemical and dynamical constraints somewhat conflict with an instability occurring 1-10 Myr after gas disk dispersal.  New isotopic data from comet 67P suggests that $\sim$22$\%$ of Earth's atmospheric noble gases were delivered via cometary impacts after the Earth had fully formed \citep{marty17}.  Because the giant planet instability is the most likely source of such a cometary onslaught, this seems to suggest that the instability occured after the conclusion of terrestrial planet formation.  However, considerable uncertainty remains in the interpretation of this noble gas signature.  Another potential conflict with our result is related to the modern Kuiper belt.  The classical Kuiper belt population on high inclination orbits can be explained if Neptune initially migrated in a slow, smooth fashion for at least 10 Myrs after gas disk dispersal before being interrupted by the giant planet instability \citep{nesvorny15a}.  Though such a timing matches our longest delay, pushing the instability time later leads to a conflict with constraints on Mars' accretion history \citep{Dauphas11}.  However, our understanding of the Kuiper belt's origins and Mars' formation timescale are subjects of ongoing study and continually evolving. Given this, we believe this is not enough to rule out the premise of our model, especially because it is able to replicate so many features of the inner solar system.

The Nice Model explains a number of aspects of the outer solar system.  We have shown that, if it occurred within 10 Myr of the dissipation of the gaseous disk, the instability produces inner solar system analogues that match many important observational constraints regarding the formation of Mars and the asteroid belt. In contrast, simulations lacking an instability consistently yield Mars analogs that are too massive, form too slowly, and are surrounded by over-developed asteroid belts.  By including a giant planet instability, these same simulations show a dramatic decrease in the mass and formation timescale of Mars, and adequate depletion in the asteroid belt.

\section{Future Work}

Our simulations are insufficient to study the large scale structure of the asteroid belt.  Simulations utilizing tens of thousands of smaller particles in the asteroid belt are necessary to test if this model can correctly match the orbital distribution of known asteroids.  Additionally, the large scale distribution of ``S-types'' and ``C-types'' is an important dynamical constraint which must be accounted for \citep{gradie82,bus02,demeo13}.  Moreover, as our knowledge of asteroids continues to undoubtedly expand through further missions such as Asteroid Redirect, Dawn, OSIRIS-Rex and Lucy, so will the number of constraints.

Another major shortcoming of our project is that we do not account for the collisional fragmentation of colliding bodies.  Higher velocity collisions can be erosive, rather than accretionary.  Given the highly excited orbits which are produced by the Nice Model instability in our simulations, accounting for the fragmentation of bodies is supremely important.  Furthermore, \citet{chambers13} showed that accounting for hit-and-run collisions in embryo accretion models can result in producing planets on colder orbits than when using standard integration schemes.  Consistent with most previous N-body simulations of the late stages of terrestrial planetary formation, our integrations fail to produce Mercury analogs with any reasonable consistency \citep{chambers01,obrien06,ray09a,kaibcowan15}.  Perhaps accounting for the fragments of material ejected in high velocity collisions with Venus analogs might provide insight into understanding this problem.

Finally, the capture of Mars' trojan satellites (some of which are rare, olivine-rich ``A-type'' asteroids) is undoubtedly affected by the strong excitation of orbits we see in the proto-Mars region of our simulations \citep{evans99}.  The trojans of Mars are the only such objects in the inner solar system with orbits stable over Gyr timescales.  Their unique compositions represent a potential observational constraint for N-body integrations of terrestrial planetary formation \citep{jac17}.

\section*{Acknowledgments}

M.S.C. and N.A.K. thank the National Science Foundation for support under award AST-1615975.  S.N.R. thanks the Agence Nationale pour la Recherche for support via grant ANR-13-BS05-0 0 03-0 02 (grant MOJO).  K.J.W. thanks NASA's SSERVI program (Institute of the Science of Exploration Targets) through institute grant number NNA14AB03A.  The majority of computing for this project was performed at the OU Supercomputing Center for Education and Research (OSCER) at the University of Oklahoma (OU).  Additional analyses and simulations were done using resources provided by the Open Science Grid \citep{osg1,osg2}, which is supported by the National Science Foundation award 1148698, and the U.S. Department of Energy's Office of Science.  Control simulations were managed on the Nielsen Hall Network using the HTCondor software package: https://research.cs.wisc.edu/htcondor/.  This research is part of the Blue Waters sustained-petascale computing project, which is supported by the National Science Foundation (awards OCI-0725070 and ACI-1238993) and the state of Illinois. Blue Waters is a joint effort of the University of Illinois at Urbana-Champaign and its National Center for Supercomputing Applications \citep{bw1,bw2}.

\bibliographystyle{apj}
\newcommand{\sci}{$Science$ }

\end{document}